\documentclass[12pt]{article}
\usepackage[latin9]{inputenc}
\usepackage{array}
\usepackage{verbatim}
\usepackage{refstyle}
\usepackage{booktabs}
\usepackage{units}
\usepackage{amsmath}
\usepackage{graphicx}
\usepackage{wasysym}
\usepackage[numbers,super,comma,sort&compress]{natbib}

\makeatletter

\AtBeginDocument{\providecommand\figref[1]{\ref{fig:#1}}}
\AtBeginDocument{\providecommand\subsecref[1]{\ref{subsec:#1}}}
\AtBeginDocument{\providecommand\secref[1]{\ref{sec:#1}}}
\AtBeginDocument{\providecommand\eqref[1]{\ref{eq:#1}}}
\AtBeginDocument{\providecommand\Eqref[1]{\ref{Eq:#1}}}
\AtBeginDocument{\providecommand\tabref[1]{\ref{tab:#1}}}
\providecommand{\tabularnewline}{\\}
\RS@ifundefined{subsecref}
  {\newref{subsec}{name = \RSsectxt}}
  {}
\RS@ifundefined{thmref}
  {\def\RSthmtxt{theorem~}\newref{thm}{name = \RSthmtxt}}
  {}
\RS@ifundefined{lemref}
  {\def\RSlemtxt{lemma~}\newref{lem}{name = \RSlemtxt}}
  {}

\providecommand*{\code}[1]{\texttt{#1}}

\usepackage{epsfig}
\usepackage{siunitx}
\usepackage{multirow}
\usepackage{placeins}
\usepackage{xr}

\usepackage{amsmath,amssymb}
\usepackage{graphicx}
\usepackage{color}
\usepackage{dcolumn}
\usepackage{bm}

\DeclareMathOperator\erf{erf}
\DeclareMathOperator\erfc{erfc}

\newref{fig}{name={Figure~}, names={Figures~}}
\newref{tab}{name={Table~}, names={Tables~}}
\newref{sec}{name={Section~}, names={Section~}}
\newref{subsec}{name={Section~}, names={Section~}}

\newenvironment{wileykeywords}
{\textsf{Keywords:}\hspace{\stretch{1}}}
{\hspace{\stretch{1}}\rule{1ex}{1ex}}

  \makeatletter
  \renewcommand\@biblabel[1]{#1.}
  \makeatother

\makeatother

\begin{document}
\title{Accelerating the 3D-RISM theory of molecular solvation with treecode
summation and cut-offs}
\author{Leighton Wilson\thanks{Department of Mathematics, University of Michigan, Ann Arbor, MI 48109},
Robert Krasny\thanks{Department of Mathematics, University of Michigan, Ann Arbor, MI 48109},
Tyler Luchko\thanks{Department of Physics and Astronomy, California State University,
Northridge, Los Angeles, CA 91330}}

\maketitle
\begin{wileykeywords} solvation, implicit solvent, 3D-RISM, AMBER, treecodes. \end{wileykeywords}
\begin{abstract}
The 3D reference interaction site model (3D-RISM) of molecular solvation
is a powerful tool for computing the equilibrium thermodynamics and
density distributions of solvents, such as water and co-ions, around
solute molecules. However, 3D-RISM solutions can be expensive to calculate,
especially for proteins and other large molecules where calculating
the potential energy between solute and solvent requires more than
half the computation time. To address this problem, we have developed
and implemented treecode summation for long-range interactions and
analytically corrected cut-offs for short-range interactions to accelerate
the potential energy and long-range asymptotics calculations in non-periodic
3D-RISM in the AmberTools molecular modeling suite. For the largest
single protein considered in this work, tubulin, the total computation
time was reduced by a factor of 4. In addition, parallel calculations
with these new methods scale almost linearly and the iterative solver
remains the largest impediment to parallel scaling. To demonstrate
the utility of our approach for large systems, we used 3D-RISM to
calculate the solvation thermodynamics and density distribution of
7-ring microtubule, consisting of 910 tubulin dimers, over 1.2 million
atoms.

\clearpage{}
\end{abstract}
\clearpage{}

\section{\textsf{INTRODUCTION}}

\begin{figure}
\begin{centering}
\includegraphics{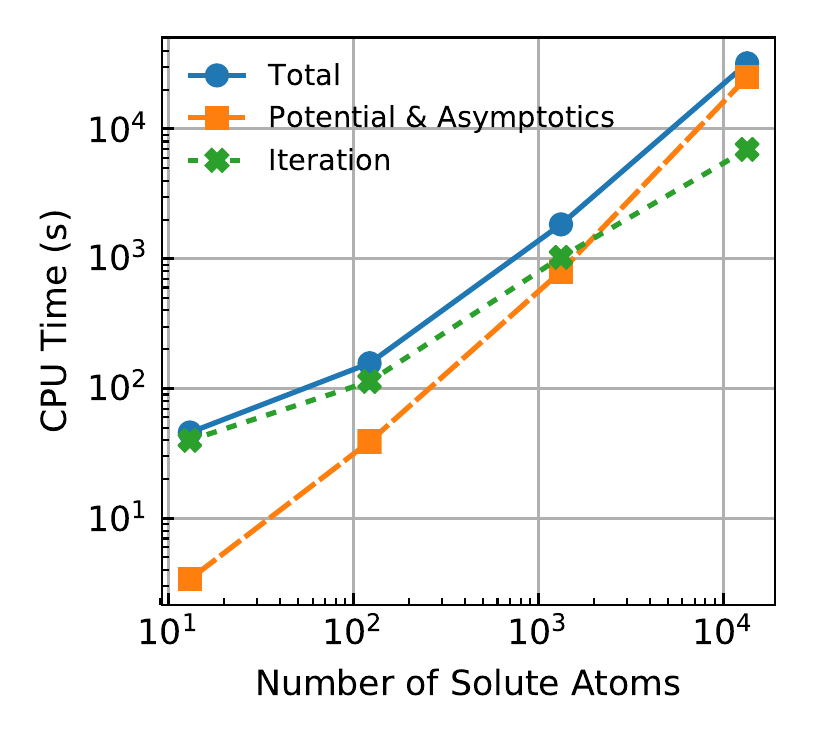}
\par\end{centering}
\caption{Time required for a single 3D-RISM calculation converged to a tolerance
of $10^{-6}$ employing direct summation. At approximately 2000 solute
atoms, the time to initialize the calculation (potential and asymptotics)
becomes larger than the time required to iterate to a solution.\label{fig:Time-required-direct-sum}}
\end{figure}

Solvation thermodynamics and the structure of the surrounding liquid
play an important role in determining the properties and interactions
of molecular systems in solution. While explicit solvent approaches
are commonly used, they can be computationally expensive and require
elaborate protocols to calculate different physical properties where
solvation is involved, such as solvation free energies \citep{duarteramosmatos2017approaches},
preferential interaction parameters \citep{giambacsu2014ioncounting},
and binding free energy\citep{cournia2017relative}. Various implicit
solvent methods have been developed to simplify and accelerate the
treatment of solvent. Among the most promising are integral equation
theories, based on the Ornstein-Zernike equation \citep{ornstein1914accidental},
and closely related classical density functional theories \citep{evans1979thenature,liu2013asite,zhao2011molecular}
as they are complete theories, calculating approximate equilibrium
distributions of explicit models, from which all solvation thermodynamics
can be computed. The 3D-reference interaction site model of molecular
solvation (3D-RISM) \citep{beglov1997anintegral,kovalenko1999selfconsistent}
is one such integral equation, which has been coupled with classical
and quantum mechanics modeling software \citep{luchko2010threedimensional,gusarov2006selfconsistent,yoshida2006anew,kloss2008quantum,miyata2008combination}
and shown to provide solvation thermodynamics in good agreement with
experiment and explicit solvent calculations \citep{johnson2016smallmolecule,palmer2010towards,sergiievskyi2014fastcomputation,truchon2014acavity}.

However, 3D-RISM can be computationally expensive, especially for
large molecules (see \figref{Time-required-direct-sum}). 3D-RISM
calculations consist of three sequential steps: initialization (calculating
potential energy and long-range electrostatic interactions on a 3D
grid), iteration to convergence, and integration of the solvent distribution
to calculate thermodynamics. For small molecules, iteration time dominates
the calculation, which scales with the number of grid points, $N_{\text{grid}}$,
as $O\left(N_{\text{grid}}\log N_{\text{grid}}\right)$. Initialization
time dominates for typical proteins, scaling with both the number
of solute atoms, $N_{\text{atom}}$, and grid points as $O\left(N_{\text{atom}}N_{\text{grid}}\right)$.
Integrating solvent thermodynamics is typically 1\% or less of the
total computation time. Depending on the precision of the calculation,
initialization becomes the most expensive part of the calculation
for solutes of 1000 atoms or more and is a major barrier to the practical
application of 3D-RISM to large molecules.

Limited work has been done to address the computational cost of initialization.
For the case of periodic boundary conditions, Heil and Kast \citep{heil20153drism}
developed the \textquotedblleft no real-space supercell\textquotedblright{}
(NRS) method. The NRS method computes all long-range Coulomb interactions
in reciprocal space, where they could be efficiently handled, scaling
as $O\left(N_{\text{grid}}\log N_{\text{grid}}\right)$ and requiring
only a small fraction of the total calculation time. This cannot be
used for open boundary conditions, where the ``supercell'' method
is used \citep{kovalenko1999selfconsistent,kovalenko2000potentials2,kaminski2010modeling}.
Because there is no periodic structure, the entire potential energy
is calculated for a real-space grid. In addition, to capture contributions
beyond the size of the solvent box, analytic long-range asymptotic
(LRA) expressions of the solvent correlation functions must also be
computed in real- and reciprocal-space. So far, little has been done
to address the cost of computing these expressions.

In this paper, we will focus on the open boundary case and address
the cost of computing potential energy and LRA functions. The potential
energy is composed of Coulomb and Lennard-Jones contributions, while
the LRA is a modified Coulomb potential. The long-range and short-range
components of these calculations require different approaches. For
short range calculations, cut-offs are often appropriate. Lennard-Jones
calculations are one such case and analytic corrections are well known
\citep{shirts2007accurate}. Long-range calculations require a different
approach.

A promising approach for dealing with long-range Coulomb-like potentials
is treecode summation. Originally developed for use in gravitational
$N$-body simulations \citep{Barnes:1986aa}, in which the force of
$N$ point masses on each other must be computed at each time step
of a dynamics simulation, they can be applied to a wide variety of
problems involving the interaction between $N$ source particles on
$M$ target sites which may or may not be coincident with the sources.
For the case of coincident sources and targets, direct summation is
$O(N^{2})$, while treecode methods are $O(N\log N)$. The essential
idea behind these methods is the replacement of particle-particle
interactions with suitably chosen particle-cluster interactions, which
can be computed approximately when the cluster and particle are well-separated.
This requires the construction of a hierarchical tree of particle
clusters and a criterion for determining when a particle and a cluster
are well-separated.

The simplest interaction between a particle and cluster is a monopole
approximation, in which, for a well-separated interaction, the cluster
is replaced with a single particle at the center whose charge (or
mass, or other property) is the sum of all particles contained within
the cluster. This is the strategy of the original gravitational Barnes--Hut
treecode \citep{Barnes:1986aa}. The fast multipole method improved
the accuracy by using higher order multipole expansions in terms of
spherical harmonics \citep{Greengard:1987aa}. Alternatively, the
interaction between a particle and a cluster can be approximated with
Cartesian Taylor expansions. However, since hard coded Taylor expansions
to high order can be costly, an alternative approach using recurrence
relations has been developed \citep{Duan:2001aa,Duan:2003aa,Li:2009aa}. 

In the evaluation of LRA functions in 3D-RISM, the solute is represented
by $N$ source particles and the solvent grid by $M$ target sites.
Traditional particle-cluster treecodes build a tree on the source
particles, with a computational cost that scales as $O(M\log N)$.
For $M\gg N,$ as is typically the case for the 3D-RISM solvent grid,
these methods scale poorly. In this case, it is advantageous to consider
an alternative cluster-particle treecode in which the tree is built
on the targets, with a computational cost that scales as $O(N\log M)$
\citep{Boateng:2013aa}. The present work demonstrates that cluster-particle
treecode summation with Taylor series recurrence relations provides
an effective approach for evaluating the Coulomb potential and the
Coulomb-like LRA functions within 3D-RISM.

In this paper, we detail the theoretical background of cluster-particle
treecode summation and our approach to cut-offs and test our implementation
in the \emph{AmberTools} molecular modeling suite \citep{AMBER2019}.
\subsecref[s]{3D-RISM} and \ref{subsec:Long-range-asymptotics} provide
a brief overview of 3D-RISM theory and potential energy and LRA functions.
In \subsecref{treecode} we detail the cluster-particle treecode method
and its application to Coulomb and screened Coulomb electrostatic
potentials. In \subsecref{treecode-lra} we extend the application
of cluster-particle treecode methods to the LRA potentials. Our error-tolerance
approach to cutoffs is presented in \subsecref[s]{Frequency-Cut-Off-for}
and \ref{subsec:Truncated-Lennard-Jones-Potential}. We provide details
of our benchmarking protocol in \secref{Methods}. Finally, in \secref{Results}
we present the results for serial benchmarking, along with practical
guidelines for selecting parameters and results for scaling with system
size and over parallel processes and application to computing the
solvation properties of microtubules.

\section{\textsf{THEORY}}

\subsection{\textsf{3D-RISM}\label{subsec:3D-RISM}}

As detailed descriptions of 3D-RISM can be found elsewhere \citep{hirata2004theoryof,kovalenko2004threedimensional,Luchko:2012aa},
we will briefly review the theory, highlighting the parts where Coulomb
interactions are involved. The 3D-RISM equation is given by
\begin{equation}
h_{\gamma}\left(\mathbf{r}\right)=\sum_{\alpha}\int c_{\alpha}\left(\mathbf{r}-\mathbf{r}^{'}\right)\chi_{\alpha\gamma}\left(r^{'}\right)\,d\mathbf{r}^{'},\label{eq:3D-RISM-r}
\end{equation}
where $\alpha$ and $\gamma$ indicate solvent sites (e.g., oxygen
or hydrogen in the case of water), and ${\bf r}$ is a grid point
location. Both $h_{\gamma}\left(\mathbf{r}\right)$, the total correlation
function (TCF), and $c_{\gamma}\left(\mathbf{r}\right)$, the direct
correlation function (DCF), are unknown quantities to be solved for
and are represented on 3D grids. $\chi_{\alpha\gamma}\left(r\right)$
is the solvent site-site susceptibility, which is computed in advance,
typically using extended RISM (XRISM) \citep{hirata1981anextended}
or dielectrically consistent RISM (DRISM) \citep{perkyns1992asitetextendashsite}.
To compute the convolution integral in \eqref{3D-RISM-r}, it is useful
to express the 3D-RISM equation in reciprocal space, 
\begin{equation}
\hat{h}_{\gamma}\left(\mathbf{k}\right)=\sum_{\alpha}\hat{c}_{\gamma}\left(\mathbf{k}\right)\hat{\chi}_{\alpha\gamma}\left(k\right),\label{eq:3D-RISM-k}
\end{equation}
where $\mathbf{k}$ is the wave vector in reciprocal space and $\hat{x}\left(\mathbf{k}\right)$
represents the Fourier transform of $x\left(\mathbf{r}\right)$.

As both the TCF and DCF are unknown, we require a closure relation
to solve for the TCF and DCF in \eqref{3D-RISM-r}. In this work we
use the Kovalenko-Hirata closure (KH) \citep{kovalenko2000potentials2},
\begin{align*}
h_{\gamma}\left(\mathbf{r}\right) & =\begin{cases}
\exp\left(d_{\gamma}\left(\mathbf{r}\right)\right)+1 & d_{\gamma}\left(\mathbf{r}\right)\le0\\
d_{\gamma}\left(\mathbf{r}\right)+2 & d_{\gamma}\left(\mathbf{r}\right)>0,
\end{cases}\\
d_{\gamma}\left(\mathbf{r}\right) & =-\beta u_{\gamma}\left(\mathbf{r}\right)+h_{\gamma}\left(\mathbf{r}\right)-c_{\gamma}\left(\mathbf{r}\right),
\end{align*}
where $u_{\gamma}\left(\mathbf{r}\right)$ is the potential energy
between solvent site $\gamma$ and the entire solute and $\beta=1/k_{B}T$
where $k_{B}$ is the Boltzmann constant and $T$ is the temperature.
The potential energy, which includes Coulomb and Lennard-Jones interactions,
is computed in advance on a 3D grid with the same dimension as that
of the TCF and DCF.

Once the TCF and DCF have been solved for, it is possible to compute
thermodynamic properties of the solvent. The most commonly used is
the excess chemical potential of the solute
\begin{equation}
\mu_{\text{ex}}^{\text{KH}}=k_{B}T\sum\limits _{\gamma}\rho_{\gamma}\times\int d{\bf r}\biggl[\frac{1}{2}(h_{\gamma}({\bf r}))^{2}\Theta\left(-h_{\gamma}\left(\mathbf{r}\right)\right)-\frac{1}{2}h_{\gamma}({\bf r})c_{\gamma}({\bf r})-c_{\gamma}({\bf r})\biggr]\label{eq:excess-chemical-potential}
\end{equation}
where $\Theta$ is the Heaviside function and $\rho_{\gamma}$ is
the bulk density of site $\gamma$. As 3D-RISM is treating a single
solute at infinite dilution, the excess chemical potential is also
the solvation free energy (SFE). Because the SFE is of general interest,
we will use it to quantify the accuracy of our numerical methods and
parameter choices.

\subsection{\textsf{Long-range asymptotics}\label{subsec:Long-range-asymptotics}}

The long-range asymptotic behavior of both the TCF and DCF are affected
by Coulomb interactions introduced in the potential energy. In particular,
the long-range asymptotics must be explicitly handled during the forward
and backward Fourier transforms used to compute the convolution integral
in \eqref{3D-RISM-k}. Several approaches have been developed to handle
this \citep{heil20153drism,perkyns2010protein,kovalenko2000potentials,kovalenko2000potentials2,kovalenko1999selfconsistent,Luchko:2012aa}.
Since we employ open boundary conditions, we use the approach originated
by Ng \citep{ng1974hypernetted} and Springer \citep{springer1973integral}
and extended for ionic solutions in 3D-RISM by Kovalenko and co-workers
\citep{kaminski2010modeling,kovalenko2000potentials,kovalenko1999selfconsistent}.
For any system where both solute and solvent have partial charges
on atomic sites, the long-range asymptotics of the DCF are given by
\begin{equation}
c_{\gamma}^{\left(\text{lr}\right)}\left({\bf r}\right)=-\beta\sum_{a}\frac{Q_{a}^{U}q_{\gamma}}{\left|{\bf r}-{\bf R}_{a}\right|}\erf\left(\frac{\left|{\bf r}-{\bf R}_{a}\right|}{\eta}\right)\label{eq:dcf-r}
\end{equation}
and 
\begin{equation}
c_{\gamma}^{\left(\text{lr}\right)}\left(\mathbf{k}\right)=-\frac{q_{\gamma}4\pi\beta}{k^{2}}\exp\left(-\frac{k^{2}\eta^{2}}{4}\right)\times\sum_{a}Q_{a}^{\text{U}}\exp\left(i\mathbf{k}\cdot\mathbf{R}_{a}\right)\label{eq:dcf-k}
\end{equation}
where $a$ is the solute site with position ${\bf R}_{a}$ and partial
charge $Q_{a}^{U}$, $\beta=\frac{1}{k_{b}T}$, $k_{b}$ is the Boltzmann
constant, $T$ is temperature, $\epsilon$ is the dielectric constant,
$\eta$ is a charge smearing parameter, and $q_{\gamma}$ is the partial
charge on solvent site $\gamma$. Both equations are computed on a
3D grid, as with the potential energy. Prior to transforming the DCF
into reciprocal space \eqref{dcf-r} is subtracted off. After the
forward Fourier transform has been performed, \eqref{dcf-k} is added
back to the DCF in reciprocal space and $\hat{h}_{\gamma}\left(\mathbf{k}\right)$
is computed from \eqref{3D-RISM-k}.

In the case that the solvent also contains ionic species, such as
Na\textsuperscript{+} or Cl\textsuperscript{-}, it is also necessary
to treat the long-range asymptotics of the TCF, given by
\begin{multline}
h_{j}^{\left(\text{lr}\right)}\left({\bf r}\right)=-\frac{\beta}{2\epsilon}\sum_{a}\frac{Q_{a}^{U}q_{j}}{\left|{\bf r}-{\bf R}_{a}\right|}\exp\left(\frac{\kappa_{D}^{2}\eta^{2}}{4}\right)\bigg[e^{\left(-\kappa_{D}\left|{\bf r}-{\bf R}_{a}\right|\right)}\erfc\left(\frac{\kappa_{D}\eta}{2}-\frac{\left|{\bf r}-{\bf R}_{a}\right|}{\eta}\right)\\
-e^{\left(\kappa_{D}\left|{\bf r}-{\bf R}_{a}\right|\right)}\erfc\left(\frac{\kappa_{D}\eta}{2}+\frac{\left|{\bf r}-{\bf R}_{a}\right|}{\eta}\right)\bigg]\label{eq:tcf-r}
\end{multline}
 and
\begin{equation}
h_{j}^{\left(\text{lr}\right)}\left(\mathbf{k}\right)=-\frac{q_{j}4\pi\beta}{\epsilon\left(k^{2}+\kappa_{D}^{2}\right)}\exp\left(-\frac{k^{2}\eta^{2}}{4}\right)\times\sum_{a}Q_{a}^{\text{U}}\exp\left(i\mathbf{k}\cdot\mathbf{R}_{a}\right)\label{eq:tcf-k}
\end{equation}
where $\kappa_{D}$ is the contribution to the inverse Debye length
of ionic co-solvent species $j$ with net charge $q_{j}$. After $\hat{h}_{\gamma}\left(\mathbf{k}\right)$
has been computed with \eqref{3D-RISM-k}, \eqref{tcf-k} is subtracted
off. Then \eqref{tcf-r} is added to the TCF after the backward Fourier
transform has been applied. \Eqref[s]{dcf-r} and (\ref{eq:tcf-r})
are also used when calculating thermodynamic observables to include
the long-range contributions not captured in the finite 3D grids used
to represent the TCF and DCF. For example, \eqref{excess-chemical-potential}
becomes
\begin{multline}
\mu_{\text{ex}}^{\text{KH}}=k_{B}T\sum\limits _{\gamma}\rho_{\gamma}\Biggl\{\int_{\text{grid}}d{\bf r}\biggl[\frac{1}{2}(h_{\gamma}({\bf r}))^{2}\Theta\left(-h_{\gamma}\left(\mathbf{r}\right)\right)-\frac{1}{2}(h_{\gamma}^{\left(\text{lr}\right)}({\bf r}))^{2}\Theta\left(Q^{\text{U}}q_{j}\right)\\
-\frac{1}{2}h_{\gamma}({\bf r})c_{\gamma}({\bf r})+\frac{1}{2}h_{\gamma}^{\left(\text{lr}\right)}({\bf r})c_{\gamma}^{\left(\text{lr}\right)}({\bf r})-c_{\gamma}({\bf r})\biggr]\\
+\int d{\bf r}\biggl[\frac{1}{2}(h_{\gamma}^{\left(\text{lr}\right)}({\bf r}))^{2}\Theta\left(Q^{\text{U}}q_{j}\right)-\frac{1}{2}h_{\gamma}^{\left(\text{lr}\right)}({\bf r})c_{\gamma}^{\left(\text{lr}\right)}({\bf r})\biggr]\Biggr\}.\label{eq:excess-chemical-potential-LRA}
\end{multline}
 Note that the first two integrals are over the volume of the grid
while the last integral is over all space and can be numerically computed
via 1D integrals \citep{kaminski2010modeling}.

\subsection{\textsf{Cluster-Particle Treecode for Electrostatic Interactions}\label{subsec:treecode}}

Consider a general potential function $\phi({\bf x},{\bf y})$, and
a collection of $M$ target sites ${\bf x}_{i}$ and $N$ disjoint
source particles ${\bf y}_{j}$ with charges $q_{j}.$ Then the potential
at a target site ${\bf x}_{i}$ is

\begin{equation}
V({\bf x}_{i})=\sum_{j=1}^{N}q_{j}\phi({\bf x}_{i},{\bf y}_{j}),\label{eq:pot}
\end{equation}
In 3D-RISM such expressions arise in computing the electrostatic potential
and the LRA functions, where the target sites lie on a regular grid,
and the source particles are the atomic solute sites, with $M\gg N$.
The cost of evaluating these expressions by direct summation is $O(MN)$.
In this section we describe the cluster-particle treecode we utilize
to reduce the cost to $O((M+N)\log M)$ \citep{Boateng:2013aa}.

\subsubsection{\textsf{Cluster-particle treecode algorithm}}

The treecode starts by building a hierarchical tree of clusters on
the $M$ target locations. The root cluster is the smallest rectangular
box containing the targets. The root is divided along all Cartesian
directions for which the side length of the root parallel to that
direction is within $\sqrt{2}$ of the shortest side length; this
yields 8, 4, or 2 child clusters. The child clusters are similarly
divided until the cluster contains less than $N_{0}$ targets, a user-specified
parameter. The tree has $L$ levels, where level 1 is the root cluster
and level $L$ contains the leaves. Each target site ${\bf x}_{i}$
belongs to a nested set of clusters ${\bf x}_{i}\in C_{L}\subseteq...\subseteq C_{1}$,
where cluster $C_{\ell}$ is at level $\ell$. Let $I_{\ell}$ denote
the set of source particles ${\bf y}_{j}$ with charge $q_{j}$ that
are well-separated from cluster $C_{\ell}$ but not from clusters
$C_{\ell-1},...,C_{1}$, and let $D$ denote the set of source particles
${\bf y}_{j}$ with charge $q_{j}$ that are not well-separated from
any cluster containing ${\bf x}_{i}$. Then \eqref{pot} can be rewritten
as 
\begin{equation}
V({\bf x}_{i})=\sum_{{\bf y}_{j}\in D}q_{j}\phi({\bf x}_{i},{\bf y}_{j})+\sum_{\ell=1}^{L}\sum_{{\bf y}_{j}\in I_{\ell}}q_{j}\phi({\bf x}_{i},{\bf y}_{j}),\label{eq:exppot}
\end{equation}
where the first term on the right hand side accounts for sources that
are close to ${\bf x}_{i}$ and the second term accounts for sources
that are well-separated from ${\bf x}_{i}.$ The first term is computed
by direct summation, and the second term is computed by Taylor approximation.
Expanding the second term $\phi({\bf x}_{i},{\bf y}_{j})$ about ${\bf x}_{c}^{\ell}$,
the center of cluster $\ell$, gives 
\begin{align}
\sum_{{\bf y}_{j}\in I_{\ell}}q_{j}\phi\left({\bf x}_{i},{\bf y}_{j}\right) & \approx\ \sum_{{\bf y}_{j}\in I_{\ell}}q_{j}\sum_{\|{\bf k}\|=0}^{p}\frac{1}{{\bf k}!}\partial_{{\bf x}}^{{\bf k}}\phi\left({\bf x}_{c}^{\ell},{\bf y}_{j}\right)\left({\bf x}_{i}-{\bf x}_{c}^{\ell}\right)^{{\bf k}}\nonumber \\
 & =\ \sum_{\|{\bf k}\|=0}^{p}m_{{\bf k}}\left({\bf x}_{c}^{\ell}\right)\left({\bf x}_{i}-{\bf x}_{c}^{\ell}\right)^{{\bf k}},\label{eq:taylor_approximation}
\end{align}
where $p$ is the order of the approximation, the coefficients $m_{{\bf k}}$
are 
\begin{equation}
m_{{\bf k}}\left({\bf x}_{c}^{\ell}\right)=\sum_{{\bf y}_{j}\in I_{\ell}}q_{j}(-1)^{\|{\bf k}\|}a_{{\bf k}}\left({\bf x}_{c}^{\ell},{\bf y}_{j}\right),
\end{equation}
and the Taylor coefficients $a_{{\bf k}}$ are 
\begin{equation}
a_{{\bf k}}\left({\bf x}_{c}^{\ell},{\bf y}_{j}\right)=\frac{1}{{\bf k}!}\partial_{{\bf y}}^{{\bf k}}\phi\left({\bf x}_{c}^{\ell},{\bf y}_{j}\right).\label{eq:taylor_coefficients}
\end{equation}
Note that \eqref{taylor_approximation} is a Taylor polynomial in
three dimensions, where $\|{\bf k}\|=k_{1}+k_{2}+k_{3}$, ${\bf k}!=k_{1}!k_{2}!k_{3}!$,
$\partial_{{\bf y}}^{{\bf k}}=\partial_{y_{1}}^{k_{1}}\partial_{y_{2}}^{k_{2}}\partial_{y_{3}}^{k_{3}}$,
$\left({\bf x}_{i}-{\bf x}_{c}\right)^{{\bf k}}=\left({x_{i}}_{1}-{x_{c}}_{1}\right)^{k_{1}}\left({x_{i}}_{2}-{x_{c}}_{2}\right)^{k_{2}}\left({x_{i}}_{3}-{x_{c}}_{3}\right)^{k_{3}}$,
and $1,2,3$ denote the three respective Cartesian directions. Also
note that the criterion for a target site and source cluster being
well-separated is $r/R<\theta$, where $r$ is the cluster radius,
$R=|x-{\bf y}_{j}|$ is the distance between the target cluster and
source cluster, and $\theta$ is the user-specified MAC parameter
\citep{Barnes:1986aa}.

\subsubsection{\textsf{Recurrence relations for Taylor coefficients of Coulomb potential}}

We illustrate this approach using the example of the Coulomb potential,
\begin{equation}
\phi({\bf x}_{i},{\bf y}_{j})=\frac{1}{\left|{\bf x}_{i}-{\bf y}_{j}\right|}.\label{eq:coulomb}
\end{equation}
In this case the Taylor coefficients in \eqref{taylor_coefficients}
can be calculated efficiently using the recurrence relation \citep{Duan:2001aa,Li:2009aa},
\begin{equation}
a_{{\bf k}}({\bf x},{\bf y})=\frac{1}{\left|{\bf x}-{\bf y}\right|^{2}}\left[\left(2-\frac{1}{\|{\bf k}\|}\right)\sum_{i=1}^{3}({x_{i}}-{y_{i}})a_{{\bf k}-{\bf e}_{i}}\right.-\left.\left(1-\frac{1}{\|{\bf k}\|}\right)\sum_{i=1}^{3}a_{{\bf k}-2{\bf e}_{i}}\right],\label{eq:coulombrecurrence}
\end{equation}
where ${\bf e}_{i}$ is the $i$th Cartesian basis vector, and $x_{i}$
represents the $i$th Cartesian component of ${\bf x}$. After explicitly
computing the coefficients for $\|{\bf k}\|=0,1$, the rest may be
computed using \eqref{coulombrecurrence}. Furthermore, if any index
of ${\bf k}$ is negative, then $a_{{\bf k}}=0$.

\subsection{\textsf{Cluster-Particle Treecode for Real-Space Long-Range Asymptotics}\label{subsec:treecode-lra}}

In addition to the Coulomb potential, the cluster-particle treecode
is utilized in 3D-RISM to compute the LRA total and direct correlation
functions, as described in this section.

\subsubsection{\textsf{Direct correlation function}}

Writing the asymptotic direct correlation function from \eqref{dcf-r}
in the cluster-particle form shown in \eqref{exppot} yields,
\begin{equation}
c_{\gamma}^{(\text{lr})}({\bf r}_{i})=\frac{-q_{\gamma}}{k_{b}T}\Biggl[\sum_{{\bf R}_{a}\in D}Q_{a}^{U}\phi_{c}^{(\text{{lr})}}({\bf r}_{i},{\bf R}_{a})+\sum_{l=1}^{L}\sum_{{\bf R}_{a}\in I_{l}}Q_{a}^{U}\phi_{c}^{(\text{{lr})}}({\bf r}_{i},{\bf R}_{a})\Biggr],\label{eq:dcfexppot}
\end{equation}
where the DCF interaction potential is
\begin{equation}
\phi_{c}^{\text{(lr)}}\left({\bf r}_{i},{\bf R}_{a}\right)=\frac{1}{\left|{\bf r}_{i}-{\bf R}_{a}\right|}\erf\left(\frac{\left|{\bf r}_{i}-{\bf R}_{a}\right|}{\eta}\right).\label{eq:dcfphi}
\end{equation}
Following \citep{Duan:2000aa}, the Taylor coefficients of the DCF
potential function in \eqref{dcfphi} are computed by the recurrence,
\begin{equation}
a_{{\bf k}}({\bf x},{\bf y})=\frac{1}{\left|{\bf x}-{\bf y}\right|^{2}}\Biggl[\left(2-\frac{1}{\|{\bf k}\|}\right)\sum_{i=1}^{3}({x_{i}}-{y_{i}})a_{{\bf k}-{\bf e}_{i}}-\left(1-\frac{1}{\|{\bf k}\|}\right)\sum_{i=1}^{3}a_{{\bf k}-2{\bf e}_{i}}+b_{{\bf k}}\Biggr],\label{eq:dcfrecurrence}
\end{equation}
where the $b_{{\bf k}}({\bf x},{\bf y})$ are the Taylor coefficients
of an auxiliary Gaussian function, $\exp\left(-\left|{\bf x}-{\bf y}\right|^{2}/\eta^{2}\right)$,
whose recurrence is 
\begin{equation}
b_{{\bf k}}({\bf x},{\bf y})=\frac{2}{\eta^{2}\|{\bf k}\|}\times\left(\sum_{i=1}^{3}({x_{i}}-{y_{i}})b_{{\bf k}-{\bf e}_{i}}-\sum_{i=1}^{3}b_{{\bf k}-2{\bf e}_{i}}\right).\label{eq:dcfauxiliaryrecurrence}
\end{equation}

\subsubsection{\textsf{Total correlation function}}

Similarly, writing the asymptotic total correlation function from
\eqref{tcf-r} in the cluster-particle form shown in \eqref{exppot}
yields,
\begin{equation}
h_{\gamma}^{(\text{lr})}({\bf r}_{i})=\frac{-q_{\gamma}}{2\epsilon k_{b}T}\exp\left(\frac{\kappa_{D}^{2}\eta^{2}}{4}\right)\Biggl[\sum_{{\bf R}_{a}\in D}Q_{a}^{U}\phi_{h}^{\text{({lr})}}({\bf r}_{i},{\bf R}_{a})+\sum_{l=1}^{L}\sum_{{\bf R}_{a}\in I_{l}}Q_{a}^{U}\phi_{h}^{(\text{{lr})}}({\bf r}_{i},{\bf R}_{a})\Biggr],\label{eq:tcfexppot}
\end{equation}
where the TCF interaction potential is
\begin{multline}
\phi_{h}^{\text{({lr})}}\left({\bf r}_{i},{\bf R}_{a}\right)=\frac{1}{\left|{\bf r}_{i}-{\bf R}_{a}\right|}\Bigg[e^{\left(-\kappa_{D}\left|{\bf r}_{i}-{\bf R}_{a}\right|\right)}\erfc\left(\frac{\kappa_{D}\eta}{2}-\frac{\left|{\bf r}_{i}-{\bf R}_{a}\right|}{\eta}\right)\\
-e^{\left(\kappa_{D}\left|{\bf r}_{i}-{\bf R}_{a}\right|\right)}\erfc\left(\frac{\kappa_{D}\eta}{2}+\frac{\left|{\bf r}_{i}-{\bf R}_{a}\right|}{\eta}\right)\Bigg].\label{eq:tcfpsi}
\end{multline}
The TCF potential function in \eqref{tcfpsi} has a complicated form
and computing its Taylor coefficients is a formidable task. Note however
that the Taylor expansions are only used when a source particle and
target cluster are well-separated, in other words when $\left|{\bf r}_{i}-{\bf R}_{a}\right|$
is large, and in that case we can take advantage of the asymptotic
properties of the complementary error function. Thus for large values
of $\left|{\bf r}-{\bf R}_{a}\right|$, we have 
\begin{align}
\erfc\left(\frac{\kappa_{D}\eta}{2}-\frac{\left|{\bf r}-{\bf R}_{a}\right|}{\eta}\right) & \approx2,\\
\erfc\left(\frac{\kappa_{D}\eta}{2}+\frac{\left|{\bf r}-{\bf R}_{a}\right|}{\eta}\right) & \approx0,
\end{align}
Using this observation, the TCF interaction potential in \eqref{tcfpsi}
is approximated by
\begin{equation}
\phi_{h}^{(\text{lr})}\left({\bf r}_{i},{\bf R}_{a}\right)\approx\frac{2\exp\left(-\kappa_{D}\left|{\bf r}_{i}-{\bf R}_{a}\right|\right)}{\left|{\bf r}_{i}-{\bf R}_{a}\right|}.\label{eq:tcfexp}
\end{equation}
Functionally, this is nothing more than a screened Coulomb interaction,
so following \citep{Li:2009aa}, we may use the recurrence relation
for its Taylor coefficients given in \eqref{screenedcoulombrecurrence},
\begin{multline}
a_{{\bf k}}({\bf x},{\bf y})=\frac{1}{\left|{\bf x}-{\bf y}\right|^{2}}\Bigg[\left(2-\frac{1}{\|{\bf k}\|}\right)\sum_{i=1}^{3}({x_{i}}-{y_{i}})a_{{\bf k}-{\bf e}_{i}}-\left(1-\frac{1}{\|{\bf k}\|}\right)\sum_{i=1}^{3}a_{{\bf k}-2{\bf e}_{i}}\\
+\kappa_{D}\left(\sum_{i=1}^{3}({x_{i}}-{y_{i}})b_{{\bf k}-{\bf e}_{i}}-\sum_{i=1}^{3}b_{{\bf k}-2{\bf e}_{i}}\right)\Bigg],\label{eq:screenedcoulombrecurrence}
\end{multline}
where the $b_{{\bf k}}({\bf x},{\bf y})$ are the Taylor coefficients
of an auxiliary exponential function, $2\exp\left(-\kappa_{D}\left|{\bf x}-{\bf y}\right|\right)$,
whose recurrence is 
\begin{equation}
b_{{\bf k}}({\bf x},{\bf y})=\frac{\kappa_{D}}{\|{\bf k}\|}\left(\sum_{i=1}^{3}({x_{i}}-{y_{i}})a_{{\bf k}-{\bf e}_{i}}-\sum_{i=1}^{3}a_{{\bf k}-2{\bf e}_{i}}\right).\label{eq:screenedcoulombauxiliaryrecurrence}
\end{equation}

\subsection{\textsf{Frequency Cut-Off for Reciprocal-Space Long-Range Asymptotics}\label{subsec:Frequency-Cut-Off-for}}

As with their real-space counterparts, direct calculation of the reciprocal-space
long-range asymptotics, \eqref[s]{dcf-k} and (\ref{eq:tcf-k}), requires
$O(MN)$ operations. However, \eqref[s]{dcf-k} and (\ref{eq:tcf-k})
decay rapidly with increasing $k^{2}$, allowing us to apply a cutoff
in $k^{2}$ beyond which it is reasonable to approximate the asymptotics
with 0. The upper bound of the error due to truncating the asympotics
is then given by
\begin{align}
\epsilon_{c,\text{tol}}^{\left(\text{lr}\right)} & =\max\left|c_{\gamma}^{\left(\text{lr}\right)}\left(\mathbf{k}_{\text{cut}}\right)\right|\nonumber \\
 & =\frac{q_{\gamma}4\sqrt{2}\pi\beta}{k_{\text{cut}}^{2}}\exp\left(-\frac{k_{\text{cut}}^{2}\eta^{2}}{4}\right)\sum_{a}Q_{a}^{\text{U}},\label{eq:k-space-dcf-err}
\end{align}
\begin{align*}
\epsilon_{h,\text{tol}}^{\left(\text{lr}\right)} & =\max\left|h_{j}^{\left(\text{lr}\right)}\left(\mathbf{k}_{\text{cut}}\right)\right|\\
 & =\frac{q_{j}4\sqrt{2}\pi\beta}{\epsilon\left(k_{\text{cut}}^{2}+\kappa_{D}^{2}\right)}\exp\left(-\frac{k_{\text{cut}}^{2}\eta^{2}}{4}\right)\sum_{a}Q_{a}^{\text{U}},
\end{align*}
where $\mathbf{k}_{\text{cut}}$ is the highest frequency wave vector
considered, and we have used the fact that $\left|\exp\left(i\mathbf{k}\cdot\mathbf{R}_{a}\right)\right|=\left|\cos\left(\mathbf{k}\cdot\mathbf{R}_{a}\right)+i\sin\left(\mathbf{k}\cdot\mathbf{R}_{a}\right)\right|\le\sqrt{2}$.
The user may then request arbitrary values for the error, or error
tolerance, in the long-range asymptotics, and an appropriate cut off
can be found using a numerical root-finder, such as Newton-Raphson.
An appropriate value for the allowable error will depend on the residual
tolerance that the 3D-RISM equations are solved to.

\subsection{\textsf{Truncated Lennard-Jones Potential}\label{subsec:Truncated-Lennard-Jones-Potential}}

It is also possible to truncate the Lennard-Jones (LJ) potential,
which is given by
\begin{align*}
u_{\text{\ensuremath{\gamma,a}}}^{\text{LJ}}\left(r\right) & =\frac{A_{\gamma,a}}{r^{12}}-\frac{B_{\gamma,a}}{r^{6}},
\end{align*}
where $A_{\gamma,a}$ and $B_{\gamma,a}$ are parameters for the specific
pair of interaction sites. Again, we can express the upper bound for
the magnitude of the error due to truncation as 
\[
\epsilon_{\text{tol}}^{\text{LJ}}=\left|u_{\text{\ensuremath{\gamma,a}}}^{\text{LJ}}\left(r_{\text{cut}}\right)\right|=\left|\frac{A_{\gamma,a}}{r_{\text{cut}}^{12}}-\frac{B_{\gamma,a}}{r_{\text{cut}}^{6}}\right|,
\]
where the cut-off distance, $r_{\text{cut}},$ can be determined for
a specific error tolerance. An appropriate value of $\epsilon_{\text{tol}}^{\text{LJ}}$
will depend on the residual tolerance that the 3D-RISM equation is
converged to. The cut-off distance for a particular error tolerance
depends on the LJ parameters of $a$ and $\gamma$, so, in practice,
we will determine separate cutoff distances for all pairs. The applied
LJ potential is then
\begin{equation}
u_{\gamma,a}^{\text{LJ-trunc}}\left(r\right)=\begin{cases}
u_{\text{\ensuremath{\gamma,a}}}^{\text{LJ}}\left(r\right) & r<r_{\text{cut},\gamma,a}\\
0 & r>r_{\text{cut},\gamma,a}
\end{cases}\label{eq:ulj-trunc}
\end{equation}
and the omitted part is
\begin{equation}
u_{\gamma,a}^{\text{LJ-omitted}}\left(r\right)=\begin{cases}
0 & r<r_{\text{cut},\gamma,a}\\
u_{\text{\ensuremath{\gamma,a}}}^{\text{LJ}}\left(r\right) & r>r_{\text{cut},\gamma,a}
\end{cases}.\label{eq:ulj-tail}
\end{equation}

However, even cut-offs with very small error tolerances can result
in large systematic errors for various observables. To determine the
impact on the excess chemical potential, we can take the functional
derivative of \eqref{excess-chemical-potential} with respect to the
potential energy,
\begin{multline}
\frac{\partial\mu_{\text{ex,KH}}}{\partial u\left(r\right)}=k_{B}T\sum\limits _{\gamma}\rho_{\gamma}\int d{\bf r}\\
\Biggl[\frac{h_{\gamma}({\bf r})\partial h_{\gamma}({\bf r})}{\partial u_{\gamma}\left(r\right)}\Theta\left(-h\left(r\right)\right)-\frac{\partial c_{\gamma}({\bf r})}{\partial u_{\gamma}\left(r\right)}-\frac{1}{2}\left(\frac{\partial h_{\gamma}({\bf r})}{\partial u_{\gamma}\left(r\right)}c_{\gamma}({\bf r})+\frac{\partial c_{\gamma}({\bf r})}{\partial u_{\gamma}\left(r\right)}h_{\gamma}({\bf r})\right)\Biggr].\label{eq:dmu-du}
\end{multline}
 For a discrete change, such as the truncation, we can rewrite this
as
\begin{multline*}
\Delta\mu_{\text{ex,KH}}=k_{B}T\sum\limits _{\gamma}\rho_{\gamma}\int d{\bf r}\\
\Biggl[h_{\gamma}({\bf r})\Delta h_{\gamma}({\bf r})\Theta\left(-h\left(r\right)\right)-\Delta c_{\gamma}({\bf r})-\frac{1}{2}\,\left(\Delta h_{\gamma}({\bf r})c_{\gamma}({\bf r})+\Delta c_{\gamma}({\bf r})h_{\gamma}({\bf r})\right)\Biggr].
\end{multline*}
 Now, consider $r_{\text{cut}}$ large enough that $u\left(r_{\text{cut}}\right)\ll1$.
Only the TCF and DCF near or beyond the cut-off will be affected,
so we use 
\begin{align}
\Delta c_{\gamma}\left(\mathbf{r}\right) & \approx-\beta\Delta u_{\gamma}\left(\mathbf{r}\right)\label{eq:lj-trunc-Delta-DCF}\\
 & =-\beta\left(u_{\gamma}^{\text{LJ}}-u_{\gamma}^{\text{LJ-trunc}}\right)\\
 & =-\beta u_{\gamma}^{\text{LJ-tail}}
\end{align}
 and neglect terms with $h_{\gamma}({\bf r})$ or $\Delta h_{\gamma}\left(\mathbf{r}\right)$,
as these decay much faster than the DCF (see \subsecref{Long-range-asymptotics}).
\Eqref{dmu-du} then simplifies to
\begin{align*}
\Delta\mu_{\text{ex}}^{\text{KH}} & =k_{B}T\sum\limits _{\gamma}\rho_{\gamma}\int d{\bf r}\left[-\Delta c_{\gamma}({\bf r})\right]\\
 & =-4\pi\sum_{a}\int_{r_{\text{cut}}}^{\infty}dr\,u_{a,\gamma}^{\text{LJ}}\left(r\right)r^{2}\\
 & =-\frac{4\pi}{3}\sum_{a}\left(\frac{1}{3}\frac{A_{\gamma,a}}{r_{\text{cut},\gamma,a}^{9}}-\frac{B_{\gamma,a}}{r_{\text{cut},\gamma,a}^{3}}\right),
\end{align*}
which is the same correction that may be applied to explicit solvent
calculations with truncated LJ interactions \citep{shirts2007accurate}.
Corrections like this can similarly be determined for other thermodynamic
observables.

\section{\textsf{METHODOLOGY}\label{sec:Methods}}

\subsection{\textsf{Benchmark System Preparation}}

\begin{table}
\begin{centering}
\begin{tabular*}{3.25in}{@{\extracolsep{\fill}}l>{\raggedleft}p{0.75in}>{\raggedleft}p{0.5in}}
\toprule 
Solute & Number of Atoms & Net Charge\tabularnewline
\midrule
Phenol & 13 & $0e$\tabularnewline
Cucurbit{[}7{]}uril & 122 & $0e$\tabularnewline
Adhiron & 1,324 & $-1e$\tabularnewline
Tubulin & 13,456 & $-36e$\tabularnewline
Microtubule (1 ring) & 174,681 & $-455e$\tabularnewline
Microtubule (7 ring) & 1,222,767 & $-3185e$\tabularnewline
\bottomrule
\end{tabular*}
\par\end{centering}
\caption{Solutes used in this work.\label{tab:Solutes-used-in}}
\end{table}

Four solutes were selected for benchmarking and testing, giving a
range in the number of atoms of over four orders of magnitude, from
13 to 13,456 atoms (see \tabref{Solutes-used-in}). For each solute,
the \code{tleap} program in AmberTools 17 \citep{case2017amber2017}
was used to assign the final parameters. OpenBabel \citep{oboyle2011openbabel,theopen}
was used to create the 3D structure of phenol from the SMILES string
``c1ccc(cc1)O''. The general Amber force field parameters (GAFF)
\citep{wang2004development} and AM1-BCC (AM1 with bond charge corrections)
charges \citep{jakalian2000fastefficient} were assigned using \code{antechamber}.
The 3D structure of cucurbit{[}7{]}uril (CB7), a neutral host molecule,
was obtained from the Statistical Assessment of the Modeling of Proteins
and Ligands 4 (SAMPL4) exercise data set \citep{muddana2014thesampl4}.
GAFF parameters were used with charges derived using the pyR.E.D.
server \citep{bayly1993awellbehaved,dupradeau2010thered,vanquelef2011redserver,wang2013redpython,frischgaussian}.
Adhiron (PDB ID: 4N6T) is an engineered scaffold protein \citep{tiede2014adhiron}
and was parameterized using Amber FF14SB \citep{maier2015ff14sbimproving}.
A complete crystal structure of tubulin, the main constituent protein
of microtubles, does not exist. We constructed a 3D model from PDB
IDs 1TVK and 1SA0 \citep{nettles2004thebinding,ravelli2004insight},
using Modeller \citep{vsali1993comparative} to combine the structures
and fill in residues missing from the H1-B2$\alpha$-tubulin loop
and the $\alpha$- and $\beta$-tubulin N-termini. The C-terminal
tails were not present in the crystal structures and replaced with
N-methylamide (NME) caps. Amber FF14SB was used for the amino acids,
the pyR.E.D. force field for GTP and GDP, and the MG\textsuperscript{2+}parameters
for use with SPC/E water from Li et al. \citep{li2013rational}.

Solvent density distributions and thermodynamics were computed for
microtubules consisting one to seven rings. A single equilibrated
microtubule ring from Ref. \citep{wells2010mechanical} was used to
construct microtubules of different lengths by replicating and translating
the ring an integer multiple of $\unit[82.746]{\mathring{A}}$ along
the microtubule axis, which is the unit cell length from the original
simulation. The resulting structure was parameterized using Amber
ff14SB \citep{maier2015ff14sbimproving} for protein, R.E.DD.B. \citep{dupradeau2008reddba}
for GTP and GDP and Li/Merz SPC/E 12-6 for Mg\textsuperscript{2+}
\citep{li2013rational}.

\subsection{\textsf{3D-RISM Calculations}}

All RISM calculations were performed in AmberTools 19 \citep{AMBER2019}.

The solvent was prepared for 3D-RISM using the \code{rism1d} program
and consisted of 55.2 M modified SPC/E water \citep{berendsen1987themissing,luchko2010threedimensional}
with 0.1 M NaCl using the corresponding Joung-Cheatham parameters
\citep{joung2008determination}. Dielectrically consistent RISM (DRISM)
theory \citep{perkyns1992asitetextendashsite} was used with a dielectric
constant of 78.44 and the Kovalenko-Hirata (KH) closure \citep{kovalenko1999selfconsistent}
at a temperature of 298.15 K. The solution was solved on 65536 grid
points with 0.025 Å grid spacing using the default parameters for
the modified direct inversion of the iterative subspace (MDIIS) solver
\citep{kovalenko1999solution}.

The \code{rism3d.snglpnt} program was used for all 3D-RISM calculations.
Default MDIIS settings, the KH closure, and a 0.5 Å grid spacing were
used for all calculations. No cut-off was used for electrostatic interactions.
The buffer distance between the solute and the edge of the solvent
box was either explicitly set or determined from the requested LJ
tolerance. In all cases, \code{rism3d.snglpnt} automatically increased
the buffer distance to ensure that all grid dimensions were divisible
by factors of 2, 3, 5, and 7, and that the number of $y$- and $z$-grid
points was divisible by the number of processes.

Performance and accuracy of the treecode summation was tested by performing
calculations using direct summation for all calculations or using
treecode for only one of DCF, TCF, or Coulomb calculations. The direct
sum benchmark calculations use a buffer distance of 24 Å and were
converged to a residual tolerance of $10^{-13}$. All other 3D-RISM
calculations detailed below were repeated five times to provide average
timings. A buffer distance of 24 Å and grid spacing of 0.5 Å were
selected as a compromise between precision and computational cost;
obtaining a relative numerical error of $10^{-10}$ would require
a solvent grid much too large to be considered. When using treecode
summation, all combinations of the MAC parameter $\theta$ from 0.2
to 0.7 in steps of 0.1, the Taylor series order $p$ from 2 to 20
in steps of 2, and maximum leaf size $N_{0}$ values of 60, 500, and
4000 were used. In all cases, the 3D-RISM equations were solved to
a residual tolerance of $10^{-10}$. Optimized serial and parallel
jobs were run with the settings in \tabref{Optimized-3D-RISM-settings.}.
To test the parallel scaling of treecode summation, calculations were
performed on 1, 2, 4, 8, 16, 24, 32, 48, 64, 72, and 96 processes
for all solutes.

Reciprocal-space cut-offs for the long-range asymptotics were compared
against the same benchmark calculations as were used for treecode
summation. In this case, the residual tolerance was varied from $10^{-10}$
to $10^{-3}$.

For the LJ cut-offs tests, we allowed the buffer distance to vary
based on the LJ tolerance. This is required to ensure that the cut-offs
fit within the solvent box so that the correction to excess chemical
potential can be applied. Only the two neutral systems were considered
as they allow us to isolate the LJ contribution from the electrostatics.
At even the lowest tolerances, the grid sizes were still manageable.
LJ and residual tolerances were independently varied from $10^{-10}$
to $10^{-3}$. The benchmark calculation in this case used LJ and
residual tolerances of $10^{-13}$.

Serial and parallel calculations for phenol, CB7, 4N6T, and tubulin
were run on our Linux cluster, Metropolis, which has seven nodes connected
by QDR Infiniband interconnects, each with 256 GB of memory and two
12 core Intel 2.4 GHz Xeon E5-2600 v2 (``Ivy Bridge-EP'') CPUs.
AmberTools was compiled with the Intel Fortran and C++ compilers 19.1.053
\citep{intel19} and the OpenMPI 3.1.3 MPI library \citep{gabriel2004openmpi}.
Additional parallel benchmarking was performed on the Skylake nodes
of Stampede2 at the Texas Advanced Computing Center through the Extreme
Science and Engineering Discovery Environment (XSEDE) \citep{towns2014xsedeaccelerating,wilkins-diehr2016anoverview},
which each have two 24 core Intel Xeon Platinum 8160 CPUs, 192 GB
of memory and are connected by a 100 Gb/sec Intel Omni-Path network.
In this case, the software was compiled with the Intel Fortran and
C++ compilers 17.0.4 \citep{intel17} and MVAPICH2 2.3 MPI library
\citep{mvapich}. Microtubule calculations were performed on Bridges
at the Pittsburgh Supercomputing Center through XSEDE \citep{towns2014xsedeaccelerating,wilkins-diehr2016anoverview},
using 16 to 24 cores per job. 

\begin{table*}
\begin{centering}
\begin{tabular}{ccccccccccc}
\toprule 
Solute & Tolerance & \multicolumn{2}{c}{TCF} &  & \multicolumn{2}{c}{DCF} &  & \multicolumn{2}{c}{Coulomb} & Reciprocal-\tabularnewline
\cmidrule{3-4} \cmidrule{4-4} \cmidrule{6-7} \cmidrule{7-7} \cmidrule{9-10} \cmidrule{10-10} 
 &  & MAC & Order &  & MAC & Order &  & MAC & Order & Space\tabularnewline
\midrule
Microtubules & $10^{-6}$ & 0.3 & 6 &  & 0.3 & 8 &  & 0.3 & 8 & $10^{-8}$\tabularnewline
Tubulin & $10^{-6}$ & 0.3 & 6 &  & 0.3 & 8 &  & 0.3 & 8 & $10^{-8}$\tabularnewline
Adhiron & $10^{-6}$ & 0.3 & 2 &  & 0.3 & 6 &  & 0.3 & 6 & $10^{-7}$\tabularnewline
CB7 & $10^{-6}$ & 0.3 & 2 &  & 0.3 & 6 &  & \multicolumn{2}{c}{Direct} & $10^{-7}$\tabularnewline
Phenol & $10^{-6}$ & 0.3 & 2 &  & \multicolumn{2}{c}{Direct} &  & \multicolumn{2}{c}{Direct} & $10^{-7}$\tabularnewline
\bottomrule
\end{tabular}
\par\end{centering}
\caption{Optimized 3D-RISM parameter settings. Treecode parameters MAC, order
$p$. All LJ cutoffs were adjusted to fit inside the solvation box.\label{tab:Optimized-3D-RISM-settings.}}
\end{table*}

\section{\textsf{RESULTS AND DISCUSSION}\label{sec:Results}}

\subsection{\textsf{Numerical precision requirements}\label{subsec:Numerical-precision-requirements}}

\begin{figure*}
\begin{centering}
\includegraphics{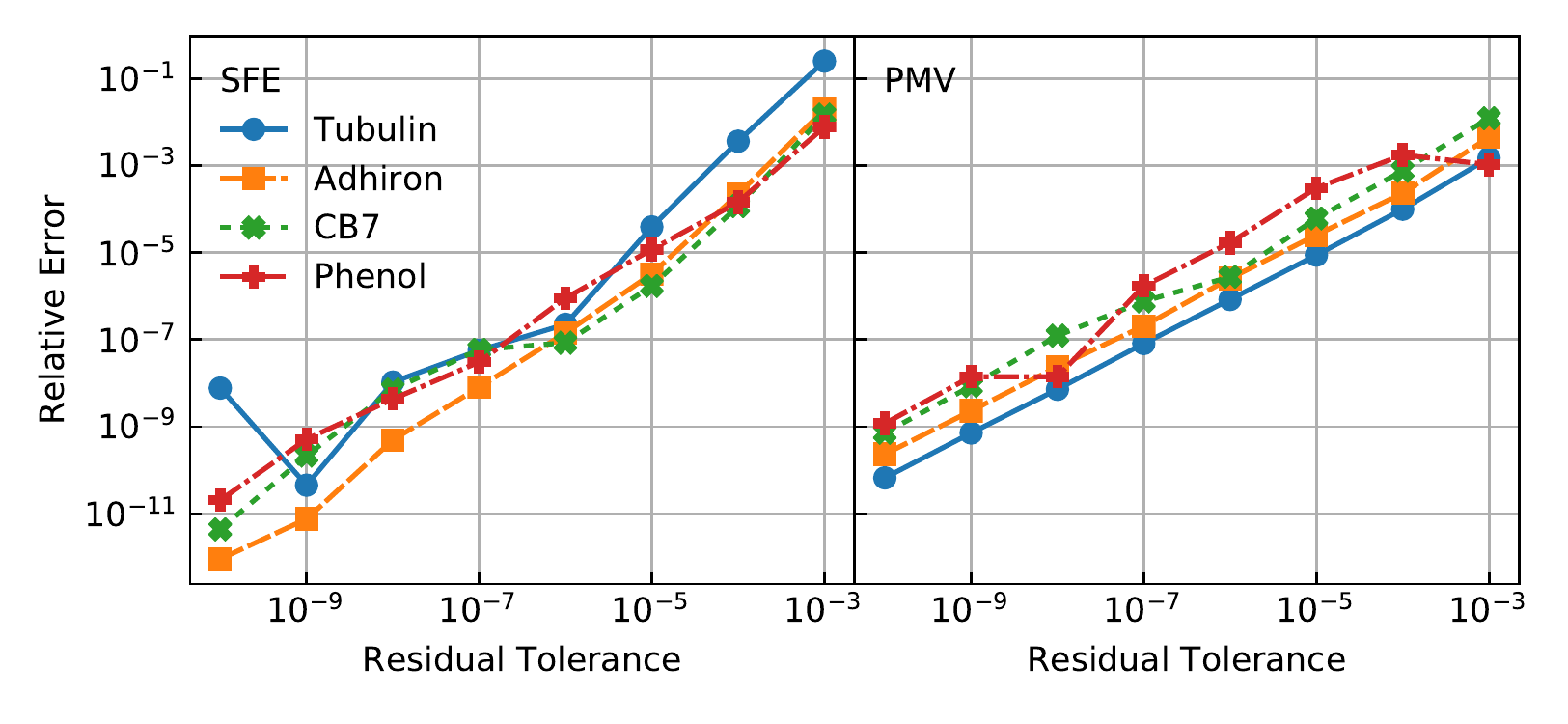}
\par\end{centering}
\caption{Dependence of the relative numerical error of the solvation free energy
(SFE) and partial molar volume (PMV) on the 3D-RISM residual tolerance.
Relative errors are calculated against a reference calculation converged
to a residual tolerance of $10^{-13}$.\label{fig:precision-vs-tolerance}}
\end{figure*}

Computational efficiencies from treecode summation and cut-offs must
not come at the cost of the numerical precision of computed thermodynamic
observables. Generally, the SFE will be the most important value to
be calculated with 3D-RISM. The numerical precision required depends
on the application to be considered. For SFE calculations absolute
errors up to $\unit[0.1]{kcal/mol}$ are generally acceptable. An
absolute error $<\unit[0.1]{kcal/mol}$ typically means relative errors
as large as $10^{-3}$ for small molecules but may need to be less
than $10^{-5}$ or even $10^{-6}$ for large proteins. To ensure stability,
molecular dynamics simulations with 3D-RISM require relative errors
less than $10^{-5}$ to ensure sufficient agreement between SFEs and
their derivatives \citep{luchko2010threedimensional}. Energy minimization
is even more demanding, requiring relative errors less than $10^{-10}$.

In practice, the convergence criterion for our iterative solver is
to reach a given maximum allowable residual tolerance. \figref{precision-vs-tolerance}
shows the relative error of SFE and PMV thermodynamic quantities as
the residual tolerance of the 3D-RISM calculation is adjusted. Overall,
we find that that residual tolerance and relative error are directly
proportional for observables we have considered. In general, we can
say that 
\begin{equation}
\epsilon_{\text{SFE}}\apprge10\times\text{tolerance}.\label{eq:error-tolerance}
\end{equation}
 For the SFE, there is no apparent dependence on the size of the solute,
though tubulin has an anomalously large relative error for a residual
tolerance of $10^{-10}$.  There does appear to be a dependence on
the solute size for the PMV, with larger solutes achieving smaller
relative errors for the same residual tolerance. The vast majority
of 3D-RISM calculations should use a residual tolerance of $10^{-5}$
or $10^{-6}$.

\subsection{\textsf{Treecode summation}}

\begin{figure*}
\begin{centering}
\includegraphics{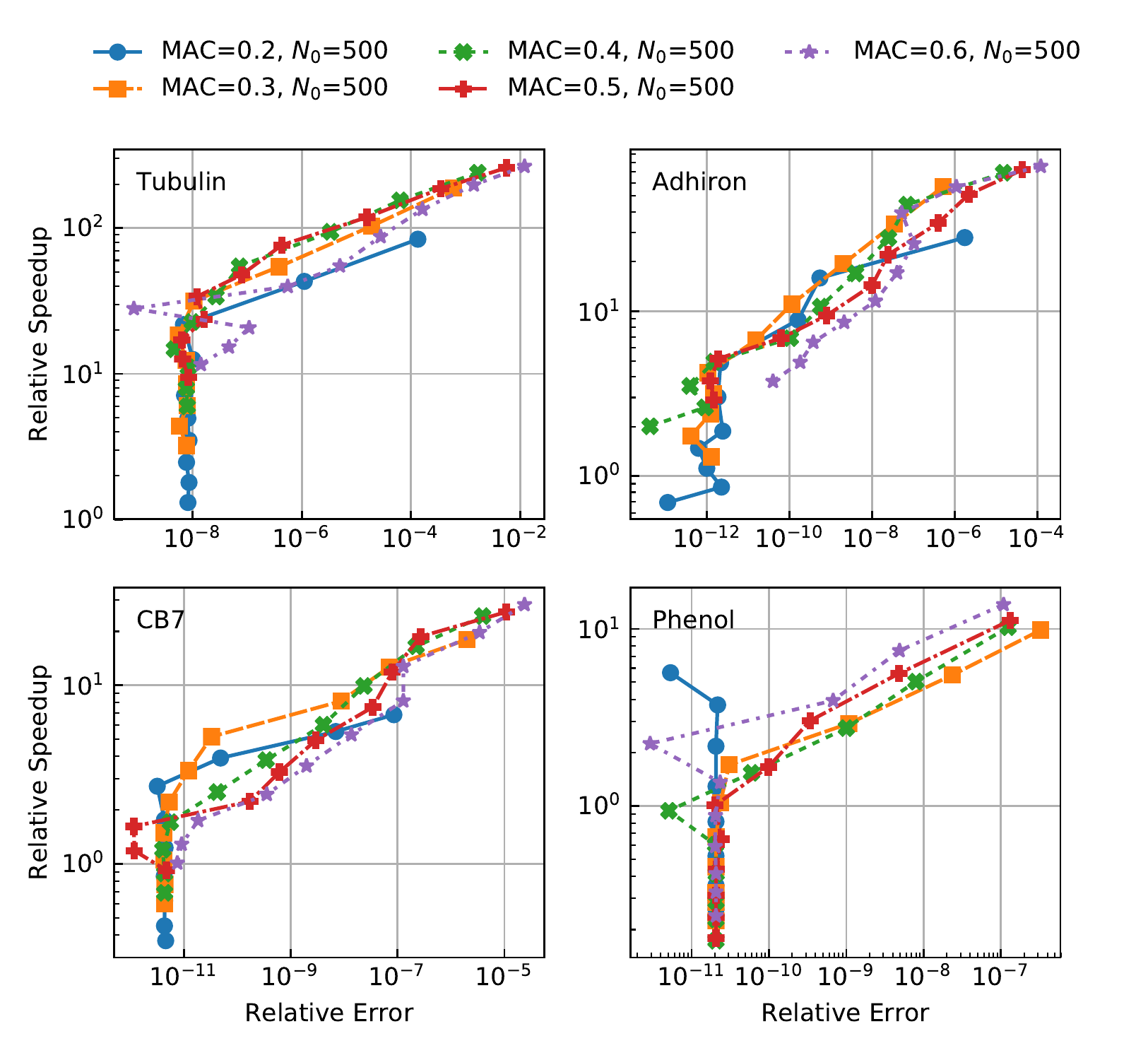}
\par\end{centering}
\centering{}\caption{Relative speedup of treecode TCF LRA compared to direct summation
vs. relative error in $\mu_{\text{ex,kh}}$ for tubulin, adhiron,
CB7, and phenol. Taylor series order $p=2k,\,k=1,\dots,10$, increasing
from right to left for each line. \label{fig:Treecode-TCF}}
\end{figure*}

\begin{figure*}
\begin{centering}
\includegraphics{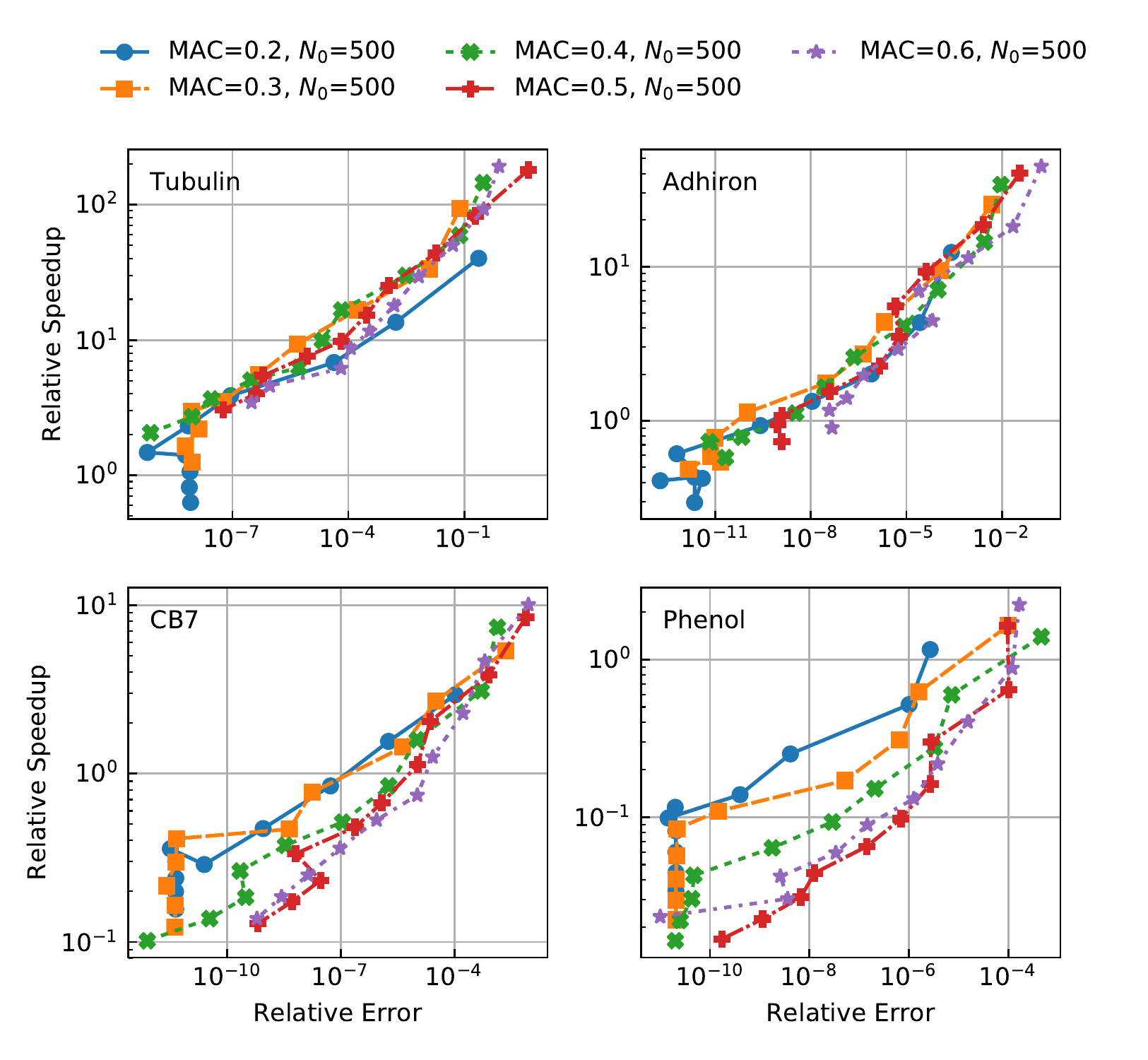}
\par\end{centering}
\centering{}\caption{Relative speedup of treecode DCF LRA compared to direct summation
vs. relative error in $\mu_{\text{ex,kh}}$ for tubulin, adhiron,
CB7, and phenol. Taylor series order $p=2k,\,k=1,\dots,10$, increasing
from right to left for each line. \label{fig:Treecode-DCF}}
\end{figure*}

\begin{figure*}
\begin{centering}
\includegraphics{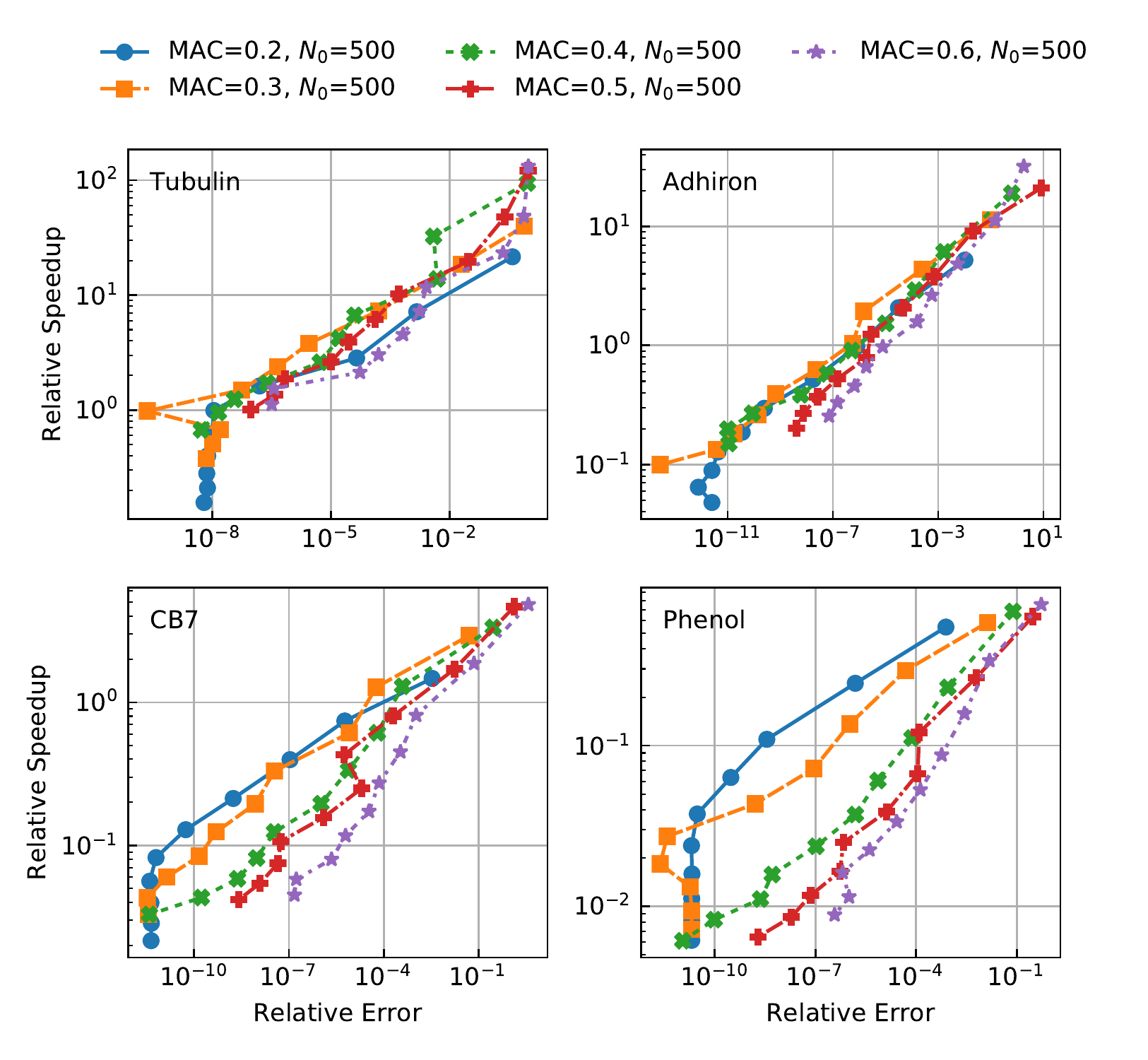}
\par\end{centering}
\centering{}\caption{Relative speedup of treecode Coulomb potential energy compared to
direct summation vs. relative error in $\mu_{\text{ex,kh}}$ for tubulin,
adhiron, CB7, and phenol. Taylor series order $p=2k,\,k=1,\dots,10$,
increasing from right to left for each line. \label{fig:Treecode-Coulomb}}
\end{figure*}

To determine the impact on speed and numerical precision of the treecode
parameters MAC, $p$, and $N_{0}$ for TCF LRA, DCF LRA, or Coulomb
potential energy, SFEs calculated from 3D-RISM for different size
solutes with different treecode parameters were compared against direct
sum calculations (\figref[s]{Treecode-TCF}, \ref{fig:Treecode-DCF}
and \ref{fig:Treecode-Coulomb}). Each data point in the plots represents
a different value of $p$ for a given MAC, increasing from right to
left. Only results for $N_{0}=500$ are shown, as we found that $N_{0}=60$
and $N_{0}=500$ performed almost identically, while $N_{0}=4000$
was generally slower for the same numerical precision. The cluster
of data points in the lower left corner of each plot indicates that
increasing $p$ does not provide any additional precision. Though
there is some noise in the timing, mostly due to interprocess interference,
increasing $p$ almost universally reduces the relative error, but
also increases execution time. In all cases, a $\text{MAC}\le0.4$
was sufficient to obtain solutions with the smallest possible error,
provided that the number of Taylor series terms was large enough.
Results for $\text{MAC}=0.7$ were omitted, as the performance was
consistently worse for all calculations. Otherwise, the best choice
of parameters depended on the quantity being summed, TCF LRA, DCF
LRA, or Coulomb potential energy, and the size of the solute.

Treecode summation shows the largest relative speedups for the TCF
LRA. In fact, treecode is faster than direct summation for all solutes
at all precisions and is nearly two orders of magnitude faster than
direct summation for tubulin and adhiron for relative errors of $10^{-5}$,
which is sufficient for most calculations. However, the treecode parameters
that give the best performance vary with the relative error and the
solute. For tubulin, $\text{MAC}=0.4$ and 0.5 have the best performance,
while $\text{MAC}=0.3$ is close. $\text{MAC}=0.3$ provides the best
performance for both adhiron and CB7, except for the largest relative
errors, where $\text{MAC}=0.4$ and even 0.5 are slightly faster.
Even phenol shows speedups relative to direct summation for all $\text{MAC}$
values with an appropriate $p$; however, the extreme values of $\text{MAC}=0.2$
and 0.6 have the best performance.

The performance of treecode summation for the DCF LRA is still much
better than direct summation for tubulin, adhiron, and CB7 but not
for phenol. In contrast to TCF LRA, it is difficult to distinguish
between the performance of different MAC values. $\text{MAC}=0.3$,
0.4 and 0.5 have similar performance for tubulin and adhiron over
almost the full range of relative errors. However, $\text{MAC}=0.5$
is unable to achieve the lowest relative errors, even for $p=20$,
and is generally slower than $\text{MAC}=0.3$ and 0.4 to achieve
the same relative error. For CB7, $\text{MAC}=0.2$ and 0.3 have nearly
identical results, outperforming larger MAC values. The trend towards
better performance from smaller MAC values continues for phenol, though
the tree code is generally slower than direct summation for this small
solute.

The Coulomb potential energy has the simplest functional form and
also shows the least benefit from treecode summation. Only tubulin
has speedups at all relative errors. However, treecode summation is
faster than direct summation for adhiron for relative errors $>10^{-7}$
and for CB7 for relative errors $>10^{-4}$. Treecode is slower than
direct summation for all phenol calculations. Otherwise, the performance
with different MAC values is similar to that observed for DCF LRA.
The best performance for tubulin and adhiron is achieved with $\text{MAC}=0.3$
and 0.4, while $\text{MAC}=0.5$ has similar performance for larger
relative errors but does not reach the lowest relative errors, even
for $p=20$. $\text{MAC}=0.2$ and 0.3 again show similar performance
for CB7, though they are faster than direct summation only for $p<4$
and $p<6$, respectively.

\subsubsection{\textsf{Treecode summation parameter selection}\label{subsec:Treecode-summation-parameter}}

\begin{figure}
\begin{centering}
\includegraphics[scale=0.8]{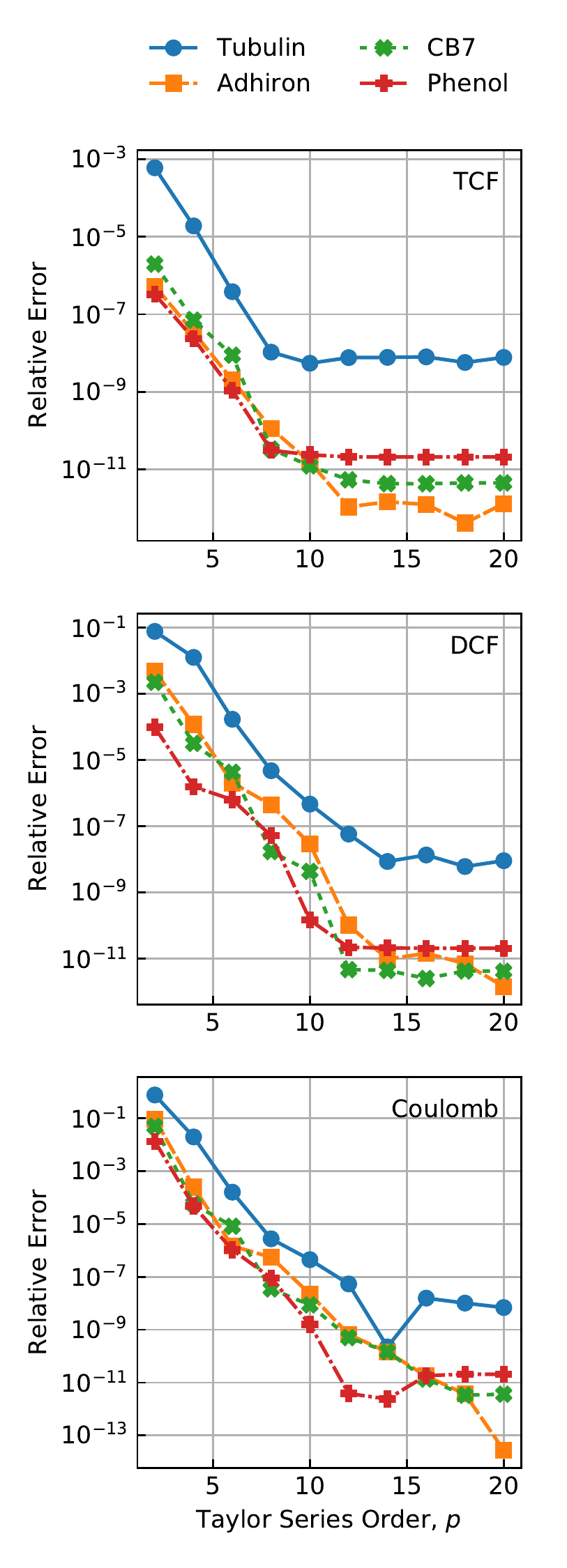}
\par\end{centering}
\caption{Relative error in $\mu_{\text{ex,kh}}$ of 3D-RISM calculations with
treecode parameters $\text{MAC}=0.3$ and $N_{0}=500$ vs. Taylor
series order, $p$, for tubulin, adhiron, CB7, and phenol. \label{fig:Treecode-all}}
\end{figure}

\begin{table*}
\centering{}%
\begin{tabular}{cccc}
\toprule 
 & MAC & $p$ & $N_{0}$\tabularnewline
\midrule
TCF & 0.3 & $\max\left(2,\frac{\log_{10}\left(\text{tolerance}\right)+5.7}{-0.7}\right)$ & 500\tabularnewline
DCF & 0.3 & $\max\left(2,\frac{\log_{10}\left(\text{tolerance}\right)+1.9}{-0.8}\right)$ & 500\tabularnewline
Coulomb & 0.3 & $\max\left(2,\frac{\log_{10}\left(\text{tolerance}\right)+1.4}{-0.8}\right)$ & 500\tabularnewline
\bottomrule
\end{tabular}\caption{Guide to selecting treecode parameters for a given residual tolerance.
Recommended parameters should be tested before production use.\label{tab:Guide-to-selecting-parameters}}
\end{table*}

Even when considering just biological molecules, there is a wide range
of shapes, sizes and charges for both the solutes and solvents that
may be studied with 3D-RISM. As a result, it is not possible to prescribe
a uniform set of parameters for treecode summation and cut-off methods
developed here; some testing will always need to be done before starting
a large calculation. However, we can provide guidance to narrow the
search for parameters that minimize computation time while preserving
necessary numerical precision. Numerical precision is set by the user
by specifying the residual tolerance at the beginning of the calculation.
As shown in \figref{precision-vs-tolerance}, relative error has a
linear relationship with the residual tolerance. Therefore, we specify
our guidelines relative to the residual tolerance.

Treecode summation requires the user to specify maximum leaf size
$N_{0}$, MAC parameter, and Taylor series order $p$. Of these, $N_{0}$
and MAC have clear best choices. $N_{0}=500$ is a safe and close
to ideal choice for all calculations; $N_{0}=60$ provides almost
identical performance while $N_{0}=4000$ gives slower performance
in some cases. If a smaller grid spacing of $\unit[0.25]{\text{Å}}$
is used, then a cluster of $N_{0}=500$ at this smaller grid spacing
will occupy about the same volume as $N_{0}=60$ for a grid spacing
of $\unit[0.5]{\text{Å}}$ and we would not expect a significant change
in performance. We also recommend $\text{MAC}=0.3$ for all calculations.
While other values can be considered for the TCF LRA calculation,
$\text{MAC}=0.3$ performs well for all calculations where treecode
is faster than direct summation. As observed for TCF and DCF LRA and
Coulomb calculations, larger MAC values perform better for larger
solutes; $\text{MAC}=0.4$ may be a better choice for solutes larger
than those considered here.

The Taylor series order is the most difficult parameter to select
as it depends on both the size of the solute, the type of calculation
being approximated, and the desired numerical precision. \figref{Treecode-all}
shows the relationship between relative error and Taylor series order
for $\text{MAC}=0.3$ and $N_{0}=500$ from \figref[s]{Treecode-TCF}
to \ref{fig:Treecode-Coulomb} grouped by calculation type across
solutes. For all solutes and all calculations, we observe a linear
relationship between $\log_{10}\left(\text{error}\right)$ and $p$
until the error due to the treecode is smaller than the error due
to reaching the convergence criterion of the iterative solver, which
is a residual tolerance of $10^{-10}$ in this case. The slope in
all cases appears similar, but there are different $y$-intercepts
for the different solutes and calculation types. In addition, tubulin
has systematically higher errors, likely due to the convergence anomaly
shown in \figref{precision-vs-tolerance}. In \tabref{Guide-to-selecting-parameters},
we provide expressions for $p$-values based on the input residual
tolerance, where we have used \eqref{error-tolerance} to relate expected
error to the input tolerance. As the case of tubulin demonstrates,
these expressions are not exact. Rather, we recommend checking the
relative error for a given $p$ by performing a test calculation with
the prescribed $p$ and another with $p+2$. If the error is sufficiently
small, then other calculations can be performed with different conformations.

It is also worth remembering that treecode summation is not always
faster than direct summation. In particular, for small molecules,
it may be better to use direct summation for the DCF LRA and Coulomb
potential.

\subsection{\textsf{Reciprocal-space cut-offs for long-range asymptotics}}

\begin{figure}
\begin{centering}
\includegraphics{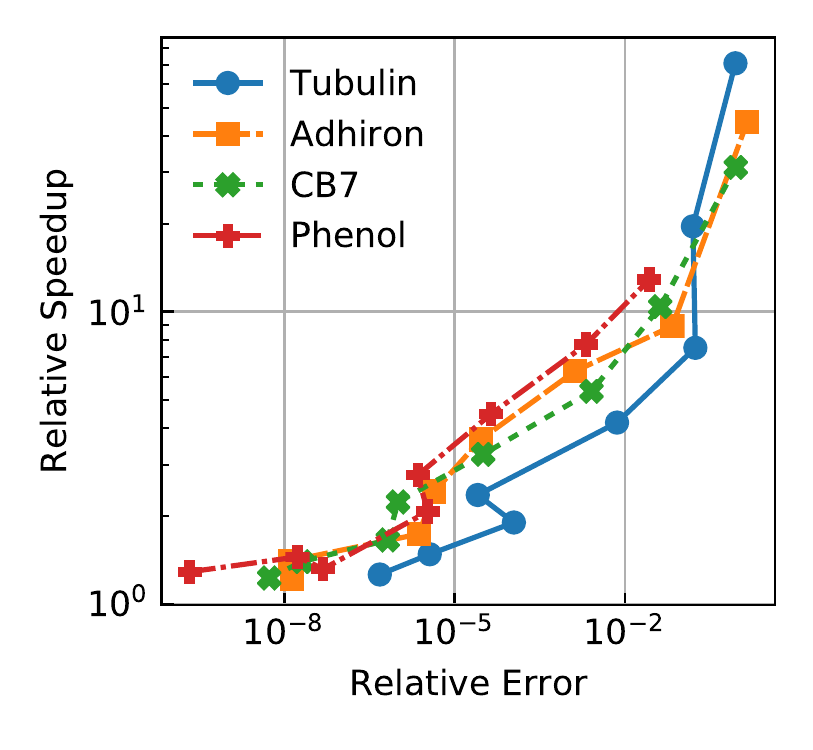}
\par\end{centering}
\caption{Relative speedup in $\mu_{\text{ex,kh}}$ vs. relative error of reciprocal
space cut-offs for TCF and DCF LRA compared to the full direct summation.
Error tolerance $\epsilon_{\text{cut}}^{\text{(lr)}}=10^{-10},\dots,10^{-3}$,
increasing from left to right for each line. \label{fig:k-space-lra-error-speedup}}
\end{figure}

Performance data for reciprocal-space LRA cut-offs is given in \figref{k-space-lra-error-speedup}.
Here, the cut-off wave number was determined from \eqref{k-space-dcf-err}
and applied to both the DCF and TCF LRA calculations. Tubulin, again,
differs from the other solutes as it requires smaller tolerances for
\eqref{k-space-dcf-err} to achieve the same relative error in the
SFE. The cutoff is based on the magnitude of the wave vector $\mathbf{k}$,
the largest value of which is determined by the grid spacing in real-space
rather than the size of the grid. In this case, we used a grid spacing
of $\unit[0.5]{\text{Å}}$, which is coarse but still of practical
use. Finer grid spacings of $0.25$ or $\unit[0.3]{\text{Å}}$ are
typically used for SFE calculations and would produce larger speed-ups
for the same cut-off. Regardless, cut-offs are always faster than
the full direct summation, even if the speedups may be small in some
cases.

\subsubsection{\textsf{Reciprocal-space cut-off parameter selection}\label{subsec:Reciprocal-space-cut-off-parameter}}

\begin{figure}
\begin{centering}
\includegraphics{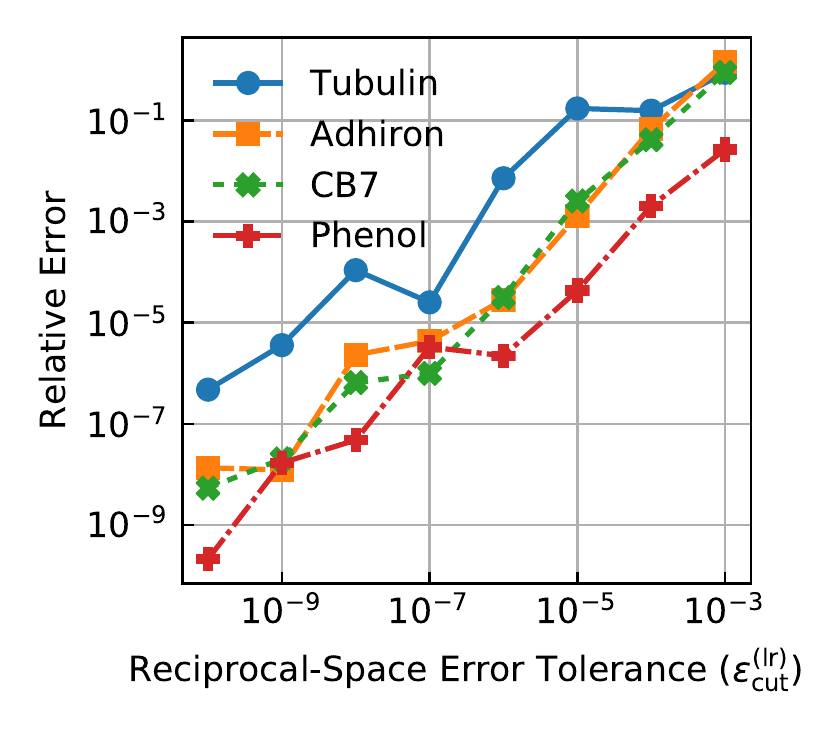}
\par\end{centering}
\caption{Relative error in $\mu_{\text{ex,kh}}$ due to applying reciprocal
space cut-offs for TCF and DCF LRA compared to full summation. \label{fig:k-space-lra-error}}
\end{figure}

To select the error tolerance for the reciprocal-space LRA cutoff,
it is useful to compare the relative error of the calculation against
the selected error tolerance, as in \figref[s]{k-space-lra-error}.
We observe, when plotted on a log-log scale, there is a nearly linear
relationship between the relative error in the SFE and the cutoff
error tolerance, suggesting that the cut-off tolerance can be selected
using the relationship
\begin{equation}
\text{error}=A\cdot\epsilon_{\text{cut}}^{B}\label{eq:error-cutoff}
\end{equation}
where $a$ and \textbf{$b$} fit to the data. Using the fact that
the relative error in the SFE is typically 10 times larger than the
specified residual tolerance, \eqref{error-tolerance}, we can rewrite
this expression as
\[
\epsilon_{\text{cut}}=a\cdot\text{tolerance}^{b}
\]
where $a=(10/A)^{1/B}$ and $b=1/B$.

From \figref{k-space-lra-error}, we observe that the reciprocal-space
LRA for tubulin incurs a larger error for the same value of $\epsilon_{\text{cut}}^{\text{(lr)}}$
compared to other solutes. Fitting to just the adhiron, CB7 and phenol
data, we arrive at 
\[
\epsilon_{\text{cut}}^{\text{(lr)}}\approx0.04\cdot\text{tolerance}^{0.9}
\]
while fitting tubulin gives
\[
\epsilon_{\text{cut}}^{\text{(lr)}}\approx0.006\cdot\text{tolerance}^{1.1}.
\]
 This roughly equates to using a cut-off tolerance that is a factor
of 10 smaller than the residual tolerance for most solutes, though
some, like tubulin, may require the cutoff tolerance to be a factor
of 100 smaller than the residual tolerance to avoid losing numerical
precision.

\subsection{\textsf{Real-space cut-offs for the Lennard-Jones potential}}

\begin{figure}
\begin{centering}
\includegraphics{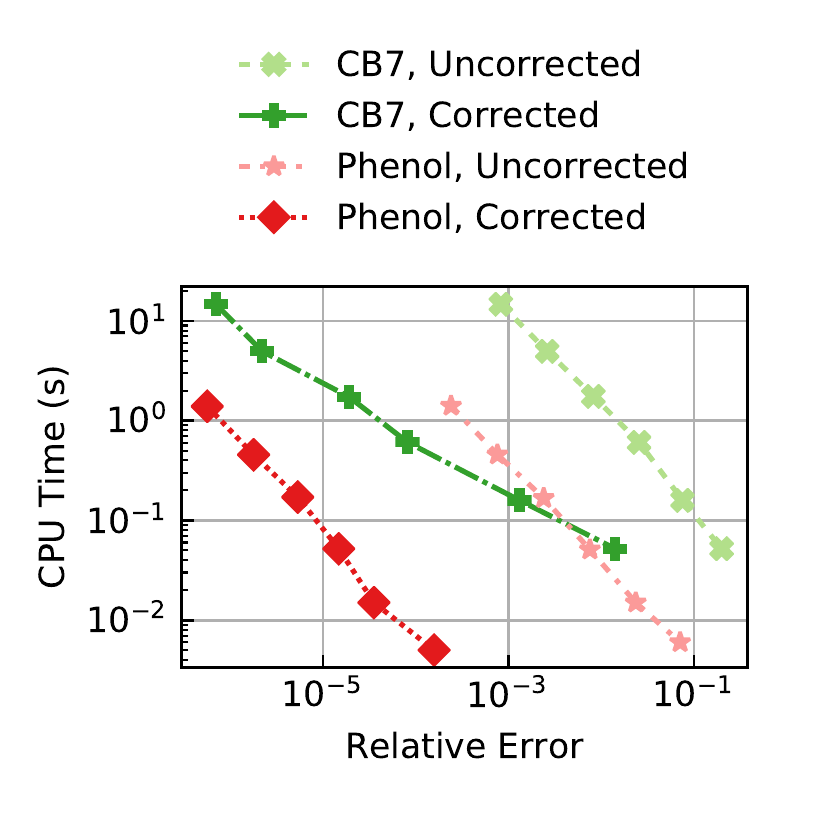}
\par\end{centering}
\caption{Calculation time for Lennard-Jones potential energy vs. SFE relative
error due to truncation with and without error correction for the
CB7 and phenol test cases. Error tolerance $\epsilon_{\text{cut}}^{\text{(LJ)}}=10^{-8},\dots,10^{-3}$,
increasing from left to right for each line.\label{fig:lj-error-time}}
\end{figure}

LJ real-space cut-offs differ from the previous calculations as we
have an analytic correction for the omitted part of the calculation.
In cases where the solute is neutral or the solvent is non-ionic,
long-range LJ iterations are the largest source of error. In these
cases, the solvation box can be safely trimmed to include the LJ cut-off
distance and nothing more. If the solute is charged and ionic solvents
or co-solvents are present, long-range electrostatics will dominate
the size of the box and the LJ calculation need only be considered
for a small part of it.

For testing purposes, we used only phenol and CB7, which are electrostatically
neutral, to clearly see the effect of the LJ cutoff and to allow us
to set the size of the solvent box to exactly accommodate the cut-off.
\figref{lj-error-time} compares the cut-off with and without the
correction where each data point from right to left reduces the LJ
error tolerance by a factor of 10 and increases the box size accordingly.
As the box-size is determined by cut-off tolerance, only the corners
are omitted from the LJ calculation; therefore, not applying cut-offs
would provide only slight differences from the uncorrected cut-offs.
While increasing the cut-off distance reduces the SFE relative error
for both CB7 and phenol, the cut-off correction cancels a significant
amount of error compared to the cut-off alone for very little computation
cost. For phenol, as the cut-off error tolerance, $\epsilon_{\text{cut}}^{\text{(LJ)}}$,
is decreased, the relative error in the SFE decreases at the same
rate for both the corrected and uncorrected data (\figref[s]{lj-error-time}
and \ref{fig:lj-error}). However, the correction reduces the relative
error by a factor of more than $10^{-2}$. As a result, the same relative
error is achieved from using the correction with $\epsilon_{\text{cut}}^{\text{(LJ)}}=10^{-3}$,
a buffer of about $\unit[10]{\text{Å}}$, as for the uncorrected calculation
with $\epsilon_{\text{cut}}^{\text{(LJ)}}=10^{-8}$, a buffer of almost
$\unit[70]{\text{Å}}$. Furthermore, because the grid size is reduced,
the corrected calculation is more than 100X faster than the uncorrected
value for both the LJ part of the calculation and the total time.
The relationship between the relative error and $\epsilon_{\text{cut}}^{\text{(LJ)}}$
is slightly different for CB7 as the correction becomes more effective
as $\epsilon_{\text{cut}}^{\text{(LJ)}}$ is lowered. Still, to achieve
the same relative error, the corrected calculation is at least 10X
faster for the same precision and is typical more than 100X faster.

\subsubsection{\textsf{Lennard-Jones cut-off parameter selection}\label{subsec:Lennard-Jones-cut-off-parameter}}

\begin{figure}
\begin{centering}
\includegraphics{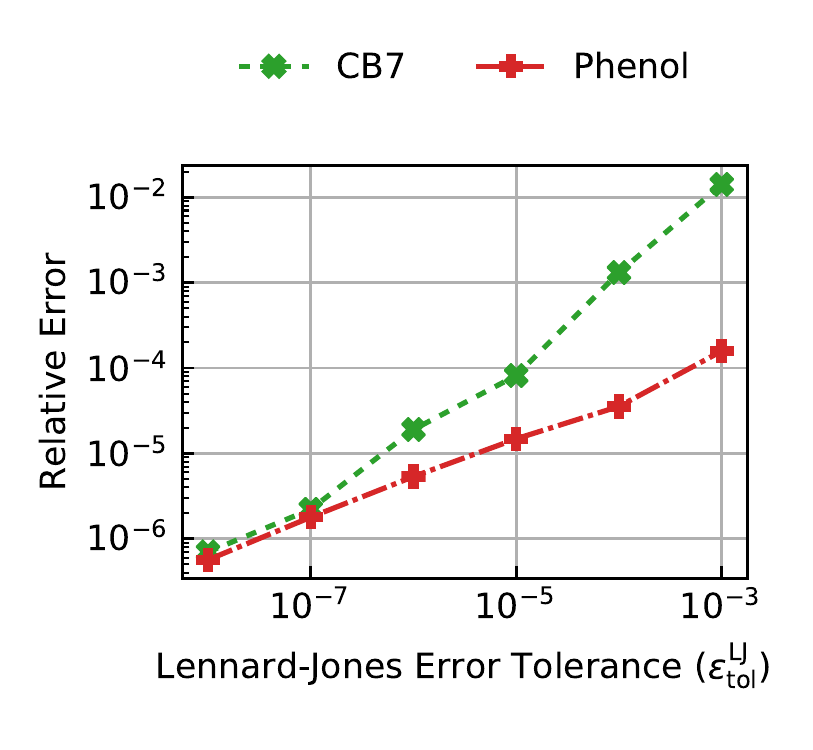}
\par\end{centering}
\caption{Relative error in $\mu_{\text{ex,kh}}$ due to applying real-space
cutoffs for the Lennard-Jones potential with different tolerances.
\label{fig:lj-error}}
\end{figure}

Only CB7 and phenol were examined for the Lennard-Jones cut-off error
tolerance and these show significantly different responses at high
values of $\epsilon_{\text{cut}}^{\text{(LJ)}}$. As with the reciprocal-space
LRA cut-off, the real-space Lennard-Jones cut-off can be selected
by comparing the relative error of the calculation against the selected
error tolerance (\ref{fig:lj-error}) and fitting with \eqref{error-cutoff}.
For phenol, we have an error estimate of 
\[
\epsilon_{\text{cut}}^{\text{(LJ)}}\approx10^{7}\cdot\text{tolerance}^{2.1}
\]
and, for CB7, we have
\[
\epsilon_{\text{cut}}^{\text{(LJ)}}\approx3\cdot\text{tolerance}^{1.1}.
\]
 The enormous difference in parameters for phenol and CB7 is due to
difference in slopes in \figref{lj-error}. A safe default choice
would be to set $\epsilon_{\text{cut}}^{\text{(LJ)}}$ to be one tenth
the residual tolerance. This should guarantee the desired precision
over a wide range of residual tolerances and, for small solutes, will
have little impact on the calculation time compared to smaller values
of $\epsilon_{\text{cut}}^{\text{(LJ)}}$. If a more aggressive optimization
is desired for larger solutes, the phenol values can be used as a
starting point and compared against calculations using larger cutoff
tolerances.

\subsection{\textsf{Scaling with solute size}}

\begin{figure*}
\begin{centering}
\includegraphics{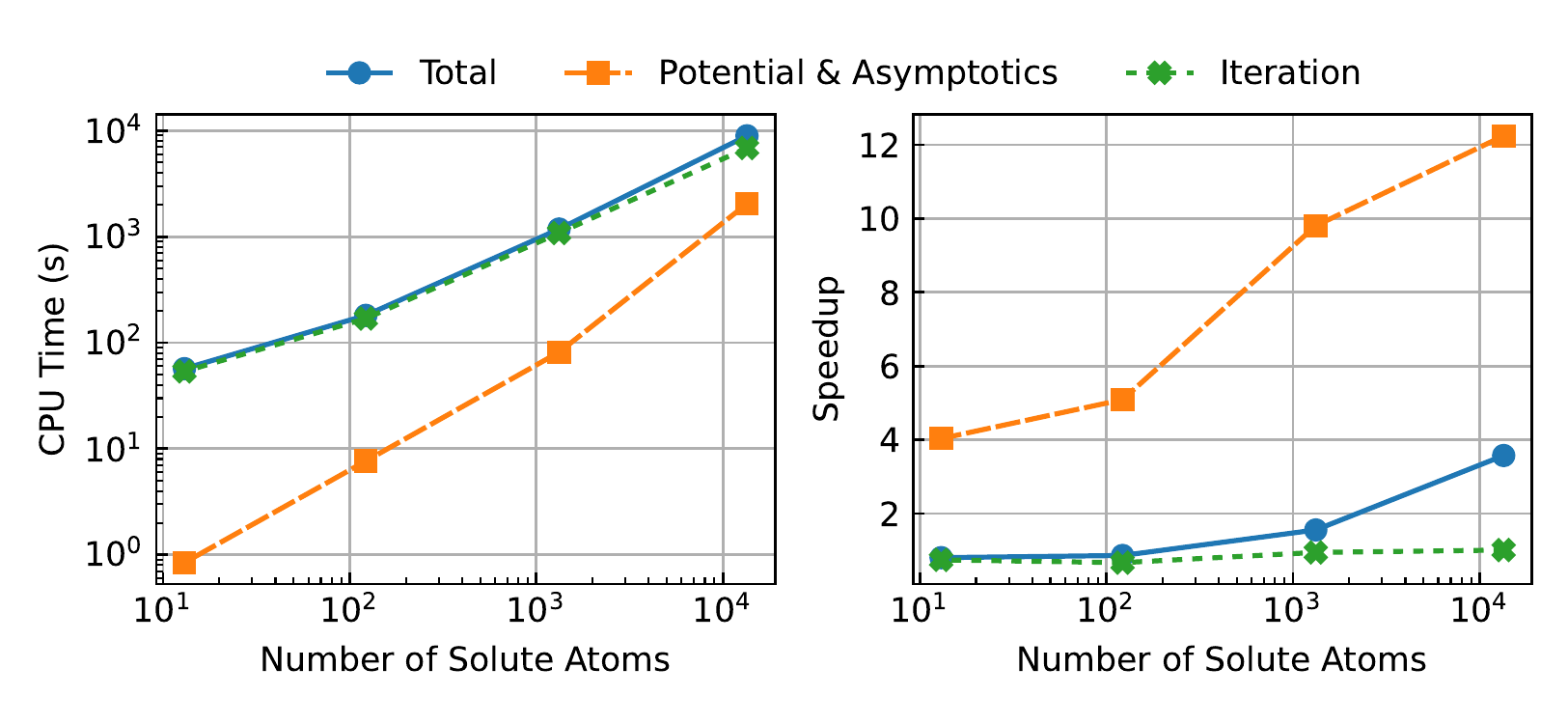}
\par\end{centering}
\caption{Total runtime of 3D-RISM converged to a tolerance of $10^{-6}$ with
potential and asymptotics calculated using treecode summation and
speedup relative to direct summation (\figref{Time-required-direct-sum}).
Required runtime is shown for setting up the calculations (potential
and asymptotics) and iterating to a converged solution. Treecode and
cut-off parameters can be found in \tabref{Optimized-3D-RISM-settings.}.
\label{fig:Total-runtime-direct-vs-tree}}
\end{figure*}

Using parameters determined in \subsecref[s]{Treecode-summation-parameter},
\ref{subsec:Reciprocal-space-cut-off-parameter} and \ref{subsec:Lennard-Jones-cut-off-parameter},
we can compare the computing time required for total cost of the calculation
with treecode summation and cut-offs with the performance of direct
summation (\figref{Total-runtime-direct-vs-tree}). For comparison
to direct summation (\figref{Time-required-direct-sum}), we again
use a residual tolerance of $10^{-6}$, which is sufficient for most
3D-RISM calculations. For larger tolerances, more aggressive parameters
can be used, resulting in potentially larger speedups. Combined, treecode
summation and cut-off methods can significantly reduce the total calculation
time -- nearly 4X faster in the case of tubulin and 1.6X for adhiron.
In the case of tubulin, computing the potential and asymptotics accounts
for about 20\% of the total runtime when using treecode and cut-offs
versus nearly 80\% using direct summation. Smaller solutes obtain
similar results; potential and asymptotics calculations are accelerated
by a factor of 3X to 10X and, with the exception of tubulin, account
for less than 10\% of the total runtime when treecode summation is
used. Overall, iteration time is now the dominant computational cost
for all solute sizes.

\begin{figure*}
\centering{}\includegraphics{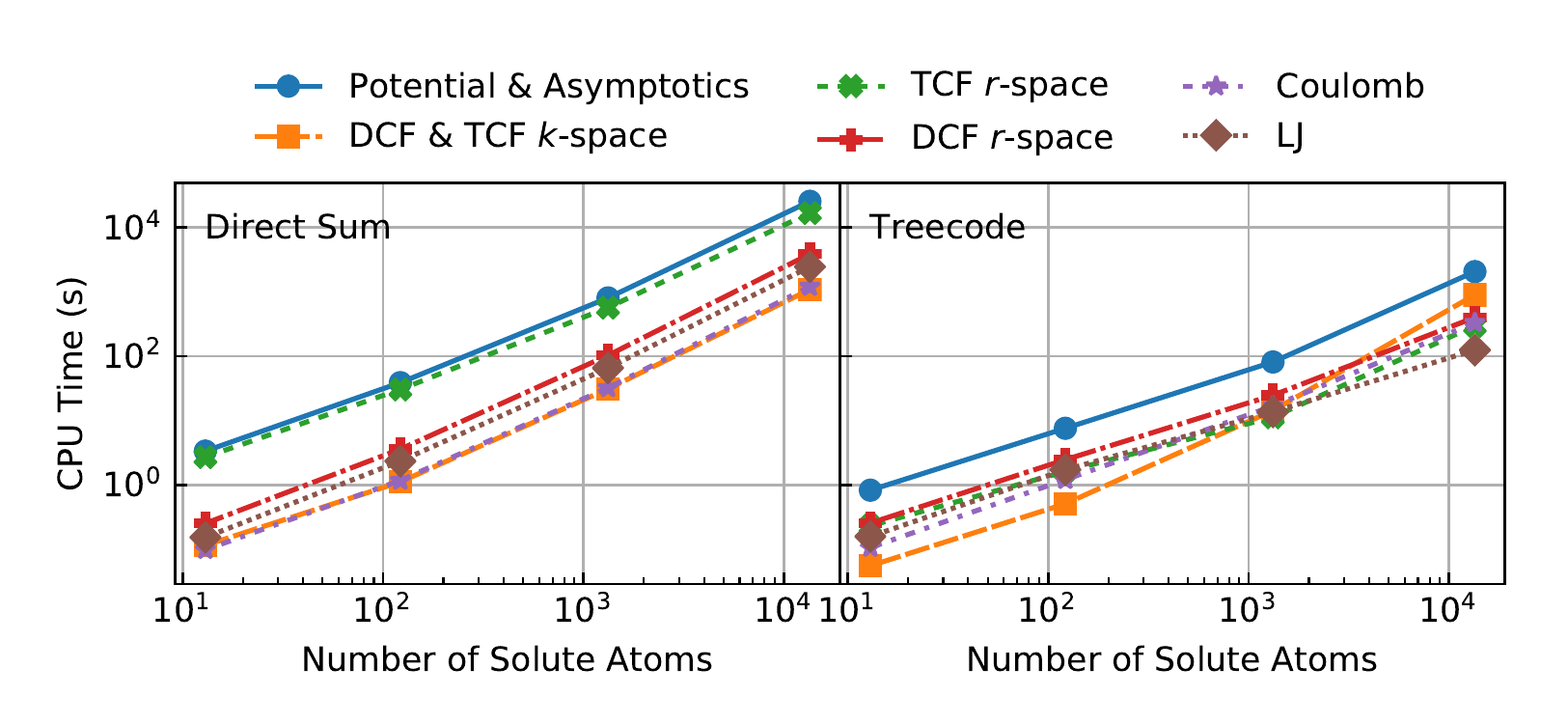}\caption{Runtime for different components of the potential and asymptotics
calculation for \figref{Total-runtime-direct-vs-tree} using direct
and treecode summation. Calculations were solved to a residual tolerance
of $10^{-6}$. \label{fig:pot.-asymp.-runtime-direct-vs-tree}}
\end{figure*}

To assess how the treecode summation and cut-off methods individually
perform, we have broken down the potential and asymptotics into their
various components (\figref{pot.-asymp.-runtime-direct-vs-tree}).
For the direct summation calculations, the real-space TCF LRA calculation
dominates the runtime, followed by the real-space DCF LRA and the
Coulomb potential energy calculations. After applying our treecode
summation and cut-off methods, the real-space DCF LRA is the most
expensive part of the calculation for all but tubulin while the real-space
TCF LRA and Coulomb potential energy require about the same amount
of time as the Lennard-Jones potential energy. Tubulin is an exception,
as the reciprocal-space DCF and TCF LRA require the largest fraction
of time, about 25\% of the total time for the potential and asymptotics.

Using a cutoff for the reciprocal-space DCF and TCF LRA is the only
optimization that does not improve the scaling with system size. As
it is a cut-off in reciprocal space, only large values of $k$ are
omitted, which are determined by the grid spacing used and have nothing
to do with solute size. As a result, we observe a performance improvement
of 2.5-3.5X for all solutes and anticipate even greater speedups for
finer grid-spacings. In fact, there should be little or no additional
computation time for calculating the reciprocal-space DCF and TCF
LRA on finer grids. Despite the fact that the scaling remains $O\left(N_{\text{atom}}N_{\text{grid}}\right)$,
the use of cut-offs means that this part of the calculation remains
a small fraction of the total and may be further reduced by other
means, such as lookup-tables.

\subsection{\textsf{Parallel scaling}}

3D-RISM in AmberTools is parallelized using the message passing interface
(MPI)  with a distributed memory model. This allows 3D-RISM to make
use of the aggregate memory of multiple nodes for large systems but
means that the code must follow the memory model of the underlying
FFT library for all of the solvation grids. We use the Fastest Fourier
Transform in the West (FFTW) \citep{FFTW3} library, which decomposes
the memory in real-space along the $z$-axis into slabs. Each process
gets one slab of each grid, whether or not that grid is directly processed
by FFTW, and includes potential energy and LRA grids. In order to
ensure adequate load balancing, 3D-RISM uses equal sized slabs for
all nodes and will automatically increase the total grid size to ensure
this if necessary. At the same time, each process gets a full copy
of the solute information. This accounts for much less memory than
the grids and is only a small fraction of the the aggregate memory
footprint, even for 96 processes.

\begin{figure*}
\begin{centering}
\includegraphics{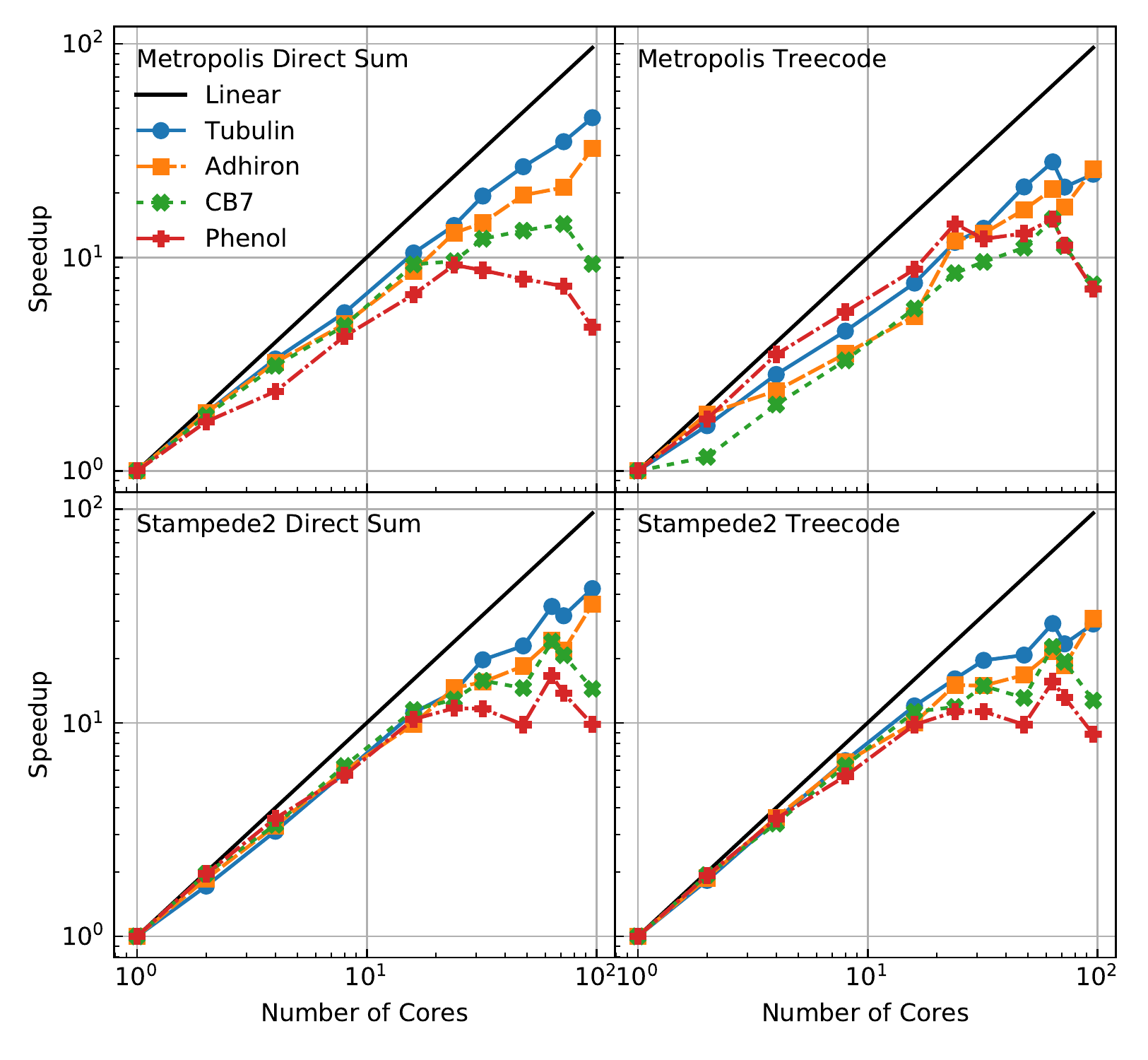}
\par\end{centering}
\caption{Speedup over multiple cores of the total calculation time for direct
and treecode summation 3D-RISM calculations converged to a tolerance
of $10^{-6}$ on Metropolis and Stampede2 clusters. Treecode and cut-off
parameters can be found in \tabref{Optimized-3D-RISM-settings.}.\label{fig:parallel-scaling}}
\end{figure*}

Treecode summation and cut-off methods have a small effect on the
overall parallel scaling of 3D-RISM (\figref{parallel-scaling}).
On the Metropolis cluster, with only 24 cores per node, calculations
on all solutes scale well until 24 cores for both types of calculations.
Adding resources beyond 24 cores causes the solution for phenol to
slow down. CB7 is the next to saturate at about 72 cores for direct
summation while adhiron and tubulin do not exhibit any slow down.
As expected, large systems scale better than smaller systems. However,
for the treecode summation and cut-off methods, 64 cores appears to
be the limit for all solutes. In addition, phenol now exhibits the
best scaling of all the systems until it saturates, while there is
a notable decline in the scaling of CB7 and adhiron.

To investigate the role of hardware, we also ran calculations on Stampede2,
which has double the cores and memory bandwidth of Metropolis (\figref{parallel-scaling}).
As with Metropolis, all solutes scale well up to 24 cores for both
direct summation and treecode/cut-off methods. Unlike Metropolis,
scaling is closer to linear and does not seem to be affected by solute
size at these small core numbers. However, 24 cores remains the scaling
limit for phenol, which indicates that this is a software limitation.
After this point, larger solutes scale more efficiently and phenol
and CB7 saturate at 24 and 64 cores respectively. Otherwise, treecode/cut-off
calculations scale as well as direct summation calculations until
the high core counts are reached.

\begin{figure*}
\begin{centering}
\includegraphics{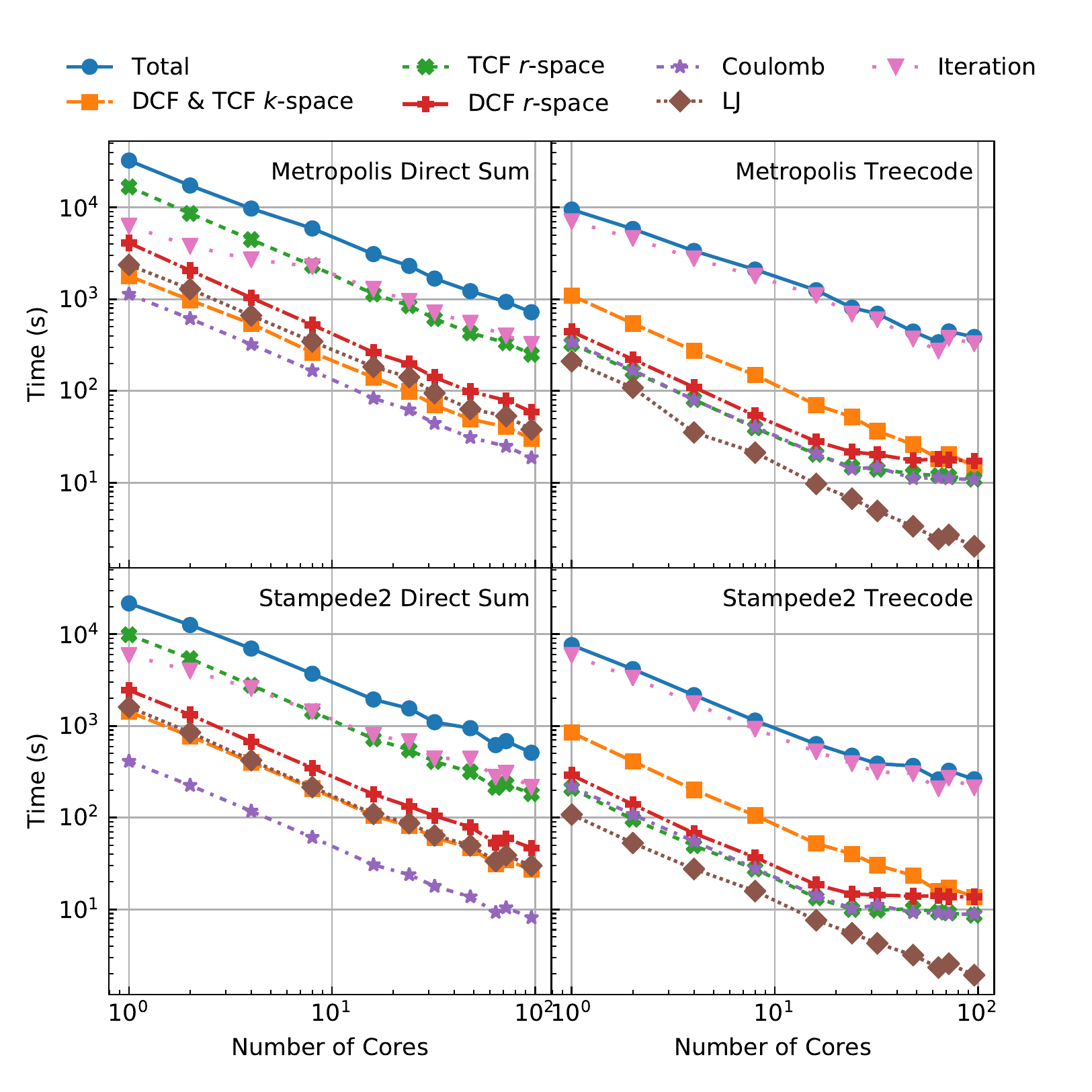}
\par\end{centering}
\caption{Computation time over multiple cores of the total calculation time
and various components for direct and treecode summation 3D-RISM calculations
on a tubulin dimer converged to a tolerance of $10^{-6}$ on Metropolis
and Stampede2 clusters. Treecode and cut-off parameters can be found
in \tabref{Optimized-3D-RISM-settings.}.\label{fig:parallel-timing-breakdown}}
\end{figure*}

As we did with single-core performance, to better understand the contributions
of different parts of the calculation, we have decomposed the calculation
into various components for the potential and asympytotics calculations
and the iteration time, the latter of which we have not attempted
to accelerate. We use tubulin for this discussion (\figref{parallel-timing-breakdown}),
though the same behavior is observed for the other molecules as well.

For direct-sum calculations, the largest bottleneck to scaling is
the iterative stage of the calculation. The scaling of this part of
the code is sub-linearly and becomes the most expensive part of the
calculation when eight or more processes are used. In contrast, all
other parts of the calculation scale almost linearly. As each MPI
process has a full copy of the solute, the direct sum calculation
is trivially parallel, with no communication between the processes,
and should scale linearly as observed. The cause of the sub-linear
scaling of the iterative calculation is beyond the scope of this paper,
but is likely hardware dependent as the iterative calculation performs
much better on Stampede2. Profiling data (not shown) indicates that
the iterative calculation has much higher memory bandwidth requirements
than the direct summation, and the higher memory bandwidth of Stampede2
could account for these differences.

Applying cut-offs to LJ and reciprocal-space calculations has little
impact on scaling. For tubulin, the selected error tolerance for the
reciprocal-space cut-offs excludes few grid points, so there is little
difference from the no-cut-off calculation. LJ cut-off calculations
do not require any communication but do limit the amount of work some
processes are required to do. However, because the cut-offs are set
to fit inside the solvation box and each solute atom has the cut-off
applied independently, the work load remains relatively balanced.
On Metropolis, the LJ cut-off calculation is slightly super-linear
but this is likely due to some small variance in the single processor
calculation. On Stampede2, scaling of LJ and reciprocal-space LRA
with and without cut-offs is nearly identical.

TCF and DCF LRA and Coulomb potential energy with treecode summation
all scale well until around 32 cores on both Metropolis and Stampede2.
The most likely reason for the scaling to plateau is that each process
performs its own treecode decomposition on its own piece of the grid.
Because a slab-decomposition memory layout is required by the FFTW3
library we use for the iterative part of the code, the memory that
each process receives becomes narrower as the process count increases.
As tree nodes narrow, it is more difficult to satisfy the MAC and
the Taylor expansion becomes less efficient. To partially alleviate
this constraint, when the tree is built, nodes are only subdivided
along a given Cartesian direction if the node box length parallel
to that direction is within a factor of $\sqrt{2}$ of the shortest
box length. However, this can result in only two or four children
in a given tree level, and the top levels of the tree will still have
node boxes with uneven aspect ratios, so narrow tree root nodes may
still affect performance. Additionally, slabs near the middle of the
grid where the solute is located may end up doing more local source
particle-target particle direct sums, while slabs near the edges of
the grid will be able to use the Taylor expansion much more often.

Overall, the performance of treecode summation for high process counts
does not adversely affect the overall parallel scaling of the calculation
as the total time and scaling is dominated by the iterative solver.
This is because treecode summation is so much faster than direct summation,
even at the highest node counts, that it is an almost negligible part
of the calculation.

\subsection{Application to microtubule stability and growth}

\begin{figure}
\centering{}\includegraphics{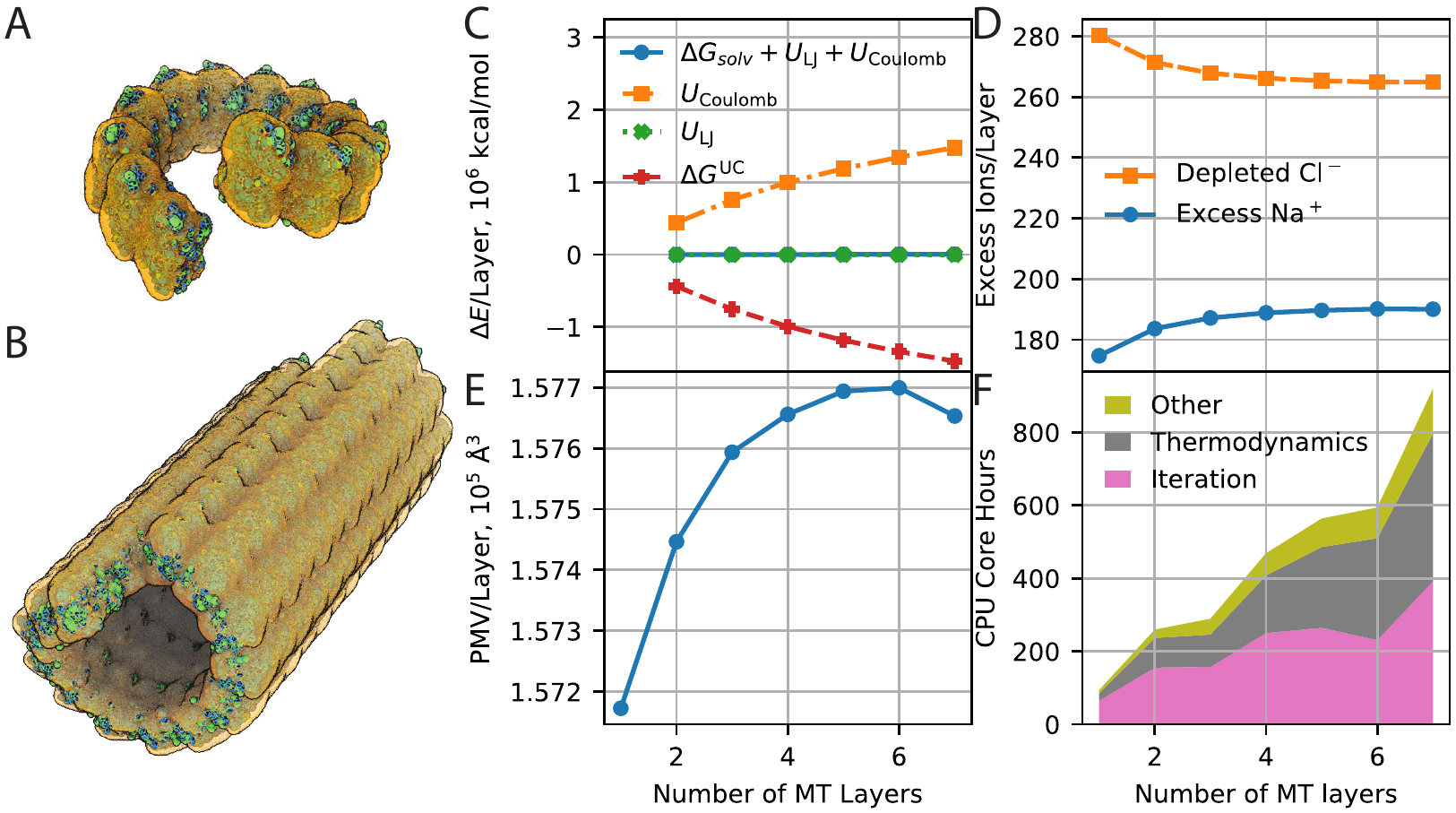}\caption{3D-RISM calculations on microtubules of various lengths. (A) A single
microtubule layer of 13 tubulin dimers and (B) a microtubule composed
of seven layers. Solvent isosurfaces are two times the bulk density
each species; transparent orange : Na\protect\textsuperscript{+},
green : Cl\protect\textsuperscript{-}, blue : water oxygen, white
: water hydrogen. Thermodynamic properties and computation time as
a function of microtubule layers: (C) Non-bonded potential energy
and solvation free energy with the Universal Correction, (D) number
of excess Na\protect\textsuperscript{+}or depleted CL\protect\textsuperscript{-}
ions per layer, (E) partial molar volume per layer, and (F) CPU core
hours for each calculation.\label{fig:MT}}
\end{figure}

Microtubules are components of the cytoskeleton found in all eukaryotic
cells and self-assemble from tubulin dimers.\citep{alberts2002molecular}
Stability is critical to the function of microtubules, particularly
during cell mitosis when they go through phases of linear growth and
rapid collapse, known as ``dynamic-instability''. Despite decades
of work, the physical mechanisms of microtubule stability and growth
are still not well understood. Here, we demonstrate how 3D-RISM can
quantify the role of ions in stabilizing microtubules. \figref{MT}
A and B shows the distribution of solvent around a single ring of
13 tubulin dimers and a microtubule composed of seven rings. It is
clear that a large excess of sodium ions surrounds the microtubule
but most of the neutralization is due to a depletion of chloride ions,
as can be seen from the preferential interaction parameter (PIP) in
molar units\citep{giambacsu2014ioncounting,smith2006equilibrium}
(\figref{MT} D), 
\[
\Gamma_{\alpha}^{\left(\text{M}\right)}=\rho_{\alpha}\int_{\text{all space}}g_{\alpha}\left(\mathbf{r}\right)-1\,d\mathbf{r}.
\]
 This is due to both the displacement of chloride and sodium ions
-- approximately 95 ion pairs per layer -- as well as the electrostatic
repulsion and attraction.  As the number of MT layers grows, the
partial molar volume of the layers remains almost constant (\figref{MT}
E), so the change in the number of excluded ions per layer is not
due to changes in the excluded volume but displacement of chloride
ions at the ring interface.

To understand the role of solvation energetically, we use the molecular
mechanics with 3D-RISM \citep{genheden2010anmm3drism} approach without
entropy to calculate the change in interaction energy and the solvation
free energy (\figref{MT} C) for adding a ring to a microtubule. Here,
the change in the effective potential energy is
\begin{align*}
\Delta E & =\Delta E_{\text{mm}}+\Delta\Delta G_{\text{solv}}\\
 & =\left(E_{\text{mm}}^{\text{MT}}-E_{\text{mm}}^{\text{MT-L}}-E_{\text{mm}}^{\text{L}}\right)\\
 & \phantom{=}\left(\Delta G_{\text{solv}}^{\text{MT}}-\Delta G_{\text{solv}}^{\text{MT-L}}-\Delta G_{\text{solv}}^{\text{L}}\right),
\end{align*}
where $\text{MT}$ and $\text{L}$ denote the final microtubule and
the added layer, $E_{\text{mm}}$ is the molecular mechanics energy,
and $\Delta G_{\text{solv}}$ is the solvation free energy, calculated
with the universal correction\citep{palmer2010towards}
\[
\Delta G_{\text{solv}}=\Delta G_{\text{3D-RISM}}+av+b,
\]
where $v$ is the partial molar volume and $a=\unit[-0.1499]{kcal/mol/\mathring{A}^{3}}$
and $b=\unit[-0.1]{kcal/mol}$ are constants fit against data from
experiment \citep{johnson2016smallmolecule}. As the rings are identical
and rigid, both the internal energy of the dimers and the entropy
are omitted and only intermolecular contributions to the energy are
included: Coulomb, Lennard-Jones and solvation solvation free energy.
 Overall, the ions in the solvent neutralize the overall charge of
the microtubule and combine with short-range inter-dimer salt bridges
to almost stabilize the structure. While not visible in \figref{MT}
C due to the scale, the change in total energy of adding one ring
slightly increases from 75 kcal/mol to 4063 kcal/mol as the microtubule
grows, while the LJ contribution is consistent at -2333 kcal/mol per
layer. The slightly positive binding energy is consistent with the
fact that we used GDP-tubulin in our calculations, which is known
to cause instability and collapse in microtubules. Since the $\Delta E$
is only about 1\% of the magnitude of $\Delta\Delta G_{\text{solv}}$
and $\Delta U_{\text{Coulomb}}$, small changes in structure, such
as an extended conformation of GTP-tubulin, could tip the balance
to stability. 

Our results show the role of the ionic environment in stabilizing
microtubules but does not elucidate the mechanism that leads to instability
when GTP is hydrolyzed to GDP. A leading hypothesis is that hydrolysis
induces conformational changes in tubulin, causing strain in the lattice
due to the kinked conformation of the GDP-tubulin or compaction of
dimers within the lattice.\citep{brouhard2018microtubule} Numerous
molecular dynamics simulations of free tubulin have failed to show
a clear conformational change between GTP- and GDP-tubulin, with both
adopting a kinked conformation, though there is evidence that GTP-tubulin
is more flexible and less strained in the microtubule lattice.\citep{igaev2020microtubule,hemmat2019multiscale,igaev2018microtubule,grafmuller2013nucleotidedependent,andre2012thestate,grafmuller2011intrinsic,bennett2009structural}
Compaction of GDP-tubulin relative to GTP-tubulin in simulation of
protofilaments has been recently reported \citep{igaev2020microtubule}
and the structure used in this calculation represents a compact lattice,
as it is composed of GDP-tubulin with a dimer repeat length of 81.2
Å \citep{wells2010mechanical}. However, microtubule lattice compaction
is not observed in all species \citep{chaaban2018thestructure}, so
it is unclear what role it plays. Our calculation could be extended
to include different conforms and isotypes to capture the role of
both inter- and intramolecular energy in microtubule stability.

Treecode summation was essential to completing these calculations,
as the total time was roughly linearly proportionate to the number
of layers added (\figref{MT} F). However, while the iterative solver
remains the most costly part of the calculation, we see, unlike our
smaller systems, that calculating the final thermodynamics now requires
a significant fraction of the total time while the potential and asymptotics
time is around 10\% of the total calculation. The reason for the increased
cost of the thermodynamics is the last integral in \eqref{excess-chemical-potential-LRA}.
Though negligible for smaller systems, it has a complexity of $O\left(N_{\text{atom}}^{2}\right)$
\citep{kaminski2010modeling} and becomes significant for systems
of this size. This will need to be addressed for practical extension
to even larger systems.

\section{\textsf{CONCLUSIONS}}

In this work, we have developed and implemented treecode summation
for long-range interactions and cut-offs for short-range interactions
to accelerate the potential and long-range asymptotics calculations
for non-periodic 3D-RISM calculations. The previous approach used
direct sum calculations that scaled as $O\left(N_{\text{grid}}N_{\text{atom}}\right)$
and were a major impediment to studying large proteins and protein
complexes. By implementing the numerical methods demonstrated here,
we have reduced the computational complexity to at most $O((N_{\text{grid}}+N_{\text{atom}})\log N_{\text{grid}})$
for almost all parts of the calculation. Furthermore, our analytic
correction for the Lennard-Jones cut-off enables much smaller solvation
boxes to be used for neutral solutes or non-ionic solvents, which
can reduce the required computation time by a factor of 100 for many
situations. 

Though proteins of almost any size will benefit from using treecode
summation and analytically corrected Lennard-Jones calculations, larger
systems benefit more. For the largest protein in our benchmark calculations,
tubulin, the total computation time was reduced by a factor of 4 and
the potential and asymptotics now account for only 20\% of the calculation
time, compared to 80\% when direct summation was used. These methods
also enabled us to calculate the solvation thermodynamics of a microtubule
composed of 910 tubulin dimers with 3D-RISM -- a calculation impossible
before now. Our results show the significant role that solvation plays
in the balance between microtubule stability and instability.

Parallel calculations with these new methods scale almost linearly
and the iterative solver remains the largest impediment to parallel
scaling. Though calculating the long-range asymptotic correction to
solvation free energy is a significant time cost for the largest microtubule
system, the iterative solver is now the most expensive part of the
calculation for almost all practical length scales and future work
will focus on accelerating this.

\section{\textsf{ACKNOWLEDGMENTS}}

TL was supported by the National Science Foundation (NSF) under Grants
CHE-1566638 and CHE-2018427 and the Research Corporation for Science
Advancement (RCSA) Cottrell Scholar Award 23967. RK and LW were supported
by NSF grant DMS-1819094 and the Michigan Institute for Computational
Discovery and Engineering. This work used the Extreme Science and
Engineering Discovery Environment (XSEDE), which is supported by National
Science Foundation grant number ACI-1548562. XSEDE Stampede 2 and
Bridges clusters at the Texas Advanced Computing Center and Pittsburgh
Supercomputing Center were used though allocation MCB190048. We thank
consultant Albert Lu for their assistance with optimizing the parallel
efficiency of 3D-RISM, which was made possible through the XSEDE Extended
Collaborative Support Service (ECSS) program. 

\bibliography{RISMTreecode}

\end{document}